\documentclass[12pt,a4paper]{article}
\pdfoutput=1
\usepackage{jcappub}

\hypersetup{colorlinks,bookmarksopen,bookmarksnumbered,citecolor=blue,
linkcolor=black,pdfstartview=FitH,urlcolor=blue}

\usepackage{graphicx}
\usepackage{amsmath}
\usepackage{amssymb}

\def\bea{\begin{eqnarray}}
\def\eea{\end{eqnarray}}
\def\ba{\begin{eqnarray}}
\def\ea{\end{eqnarray}}
\def\be{\begin{equation}}
\def\ee{\end{equation}}
\def\beq{\begin{equation}}
\def\eeq{\end{equation}}

\newcommand{\lsim}{\mathrel{\rlap{\lower4pt\hbox{\hskip1pt$\sim$}}
    \raise1pt\hbox{$<$}}}         
\newcommand{\gsim}{\mathrel{\rlap{\lower4pt\hbox{\hskip1pt$\sim$}}
    \raise1pt\hbox{$>$}}}         

\newcommand{\leftrightarrowraised}{\mathrel{\rlap{\lower-0pt\hbox{\hskip1pt$\partial$}}
    \raise6 pt\hbox{$\leftrightarrow$}}}

\newcount\hour \newcount\minute
\hour=\time \divide \hour by 60
\minute=\time
\title{Pinning down inelastic dark matter in the Sun and in direct detection}

\def\kth{Department of Theoretical physics,
School of Engineering Sciences, KTH Royal Institute of Technology,
AlbaNova University Center, 106 91 Stockholm, Sweden}

\author{\textbf{Mattias Blennow,}\vspace*{0mm}}
\author{\textbf{Stefan Clementz}\vspace*{0mm}}
\author{\textbf{and Juan Herrero-Garcia}\vspace*{0mm}}
\affiliation{\kth}

\abstract{We study the solar capture rate of inelastic dark matter with endothermic and/or exothermic interactions. By assuming that an inelastic dark matter signal will be observed in next generation direct detection experiments we can set a lower bound on the capture rate that is independent of the local dark matter density, the velocity distribution, the galactic escape velocity as well as the scattering cross section. In combination with upper limits from neutrino observatories we can place upper bounds on the annihilation channels leading to neutrinos. We find that, while endothermic scattering limits are weak in the isospin-conserving case, strong bounds may be set for exothermic interactions, in particular in the spin-dependent case. Furthermore, we study the implications of observing two direct detection signals, in which case one can halo-independently obtain the dark matter mass and the mass splitting, and disentangle the endothermic/exothermic nature of the scattering. Finally we discuss isospin violation.}

\keywords{dark matter theory, dark matter experiments}

\begin{document}
\maketitle

\section{Introduction} \label{s:intro}

There is now an overwhelming amount of evidence in favor of the existence of non-baryonic dark matter (DM), and several searches have been designed to look for non-gravitational signals. As the Earth and the Sun move through the DM halo of the Milky Way scatterings of the DM particles with nuclei are expected. Direct detection (DD) experiments are designed to look for the nuclear recoils produced inside underground detectors due to these scatterings~\cite{Lee.:2007qn,Archambault:2009sm,Bernabei:2010mq,Felizardo:2011uw,Armengaud:2012pfa,Aprile:2012nq,Aalseth:2012if,Aprile:2013doa,Li:2013fla,Agnese:2014aze,Angloher:2014myn,Akimov:2011tj,Akerib:2013tjd,Xiao:2014xyn,Amole:2015lsj,Akerib:2016lao}. The DM particles may also scatter off nuclei in the Sun and in the Earth and become gravitationally bound. They will then accumulate in the core of the body in which they are captured due to subsequent scatterings where they may annihilate when their density is large enough, producing a flux of high-energy neutrinos detectable in neutrino telescopes~\cite{Tanaka:2011uf, Aartsen:2012kia,Adrian-Martinez:2013ayv,Choi:2015ara}. These scattering rates, either in man-made detectors or in astrophysical objects, depend on the DM velocity distribution in the halo, which is typically assumed to be a Maxwell-Boltzmann distribution with a cut-off at the galactic escape velocity, known as the Standard Halo Model (SHM). However, there are large uncertainties in the velocity distribution and the local background density, which make it desirable to draw conclusions that do not depend on it. In particular, in order to check the compatibility of different signals being independent of the astrophysics is of uttermost importance.

The aim of this paper is to study the solar capture of one of the most popular scenarios for a dark sector: inelastic DM. We study both the endothermic and the exothermic cases, i.e., when the particle scatters into a heavier or a lighter mass state after the interaction. Furthermore, we analyse which properties of the DM states that can be inferred from DD signals and we generalize the halo-independent lower bound on the capture of DM in the Sun based on the observation of a DM signal in a DD experiment, derived in ref.~\cite{Blennow:2015oea}, to the case of inelastic interactions.

Inelastic DM was introduced to solve the tension between the annual modulation signal observed in DAMA~\cite{Bernabei:2010mq} and the upper limits from other experiments~\cite{TuckerSmith:2001hy}, see also refs.~\cite{TuckerSmith:2004jv,Schwetz:2011aa}. Usually the inelastic interaction is assumed to be endothermic. More recently, ideas have been put forth to alleviate the tension between DD experiments by introducing exothermic DM~\cite{Graham:2010ca,Frandsen:2014ima,Chen:2014tka}. In these models, the heavier state is also populated and in a collision downscatters to the lower mass state. In the following, we will define the mass-splitting parameter $\delta$ as
\be
\delta = m_{\chi^*}-m_{\chi}\,,
\ee
where the particle scatters from a state of mass $m_{\chi}$ to a different state of mass $m_{\chi^*}$. The process is endothermic for $\delta > 0$ and exothermic if $\delta < 0$. Of course, elastic interactions are recovered when $\delta=0$. In this paper we will be interested in splittings of $\mathcal{O}(10-100)$ keV. Due to the mass-splitting, the kinematics of inelastic DM scattering changes significantly from the elastic case, and thus also the capture rate in the Sun and the event rate in a DD experiment are drastically different.

The remainder of this paper is organized as follows: In section~\ref{s:pheno} we discuss simple scenarios of inelastic DM and the stability of the excited state, which is crucial to have either endothermic or exothermic interactions. We then discuss the kinematics of the solar capture of inelastic DM, its thermalization, and the neutrino signal due to DM annihilations in section~\ref{s:capt}. Section~\ref{s:DD} provides a discussion on the direct detection (DD) of inelastic DM with particular emphasis on the different predictions for endo- and exothermic scattering. We also study the case where 2 different signals are observed. Furthermore, we review how the halo DM velocity distribution can be extracted from the signal observed in a DD experiment in order to set a lower bound on the solar capture rate. We illustrate this bound using mock data in section~\ref{s:bound}, deriving limits on the different channels from neutrino observatories like Super-Kamiokande (SK)~\cite{Tanaka:2011uf,Choi:2015ara} and IceCube (IC)~\cite{Aartsen:2012kia}. Finally, we give our concluding remarks in section~\ref{s:conc}.


\section{Endothermic and exothermic inelastic dark matter} \label{s:pheno}

In this work we consider both endo- and exothermic DM interactions in a phenomenological fashion. However, it is useful to keep in mind the different plausible scenarios of inelastic DM, as they give rise to very different phenomenology, see e.g., ref.~\cite{Batell:2009vb}. We assume that DM is neutral under the SM gauge group and charged under an abelian symmetry, which we take to be a gauge  $\rm U(1)_{\rm X}$ with coupling constant $g_X$ for simplicity. The DM charge is $Q^{\rm X}_{\rm DM}=1$. The gauge boson which mediates the interactions is denoted by $A'_\mu$ with mass $m_{A'}$ after spontaneous symmetry breaking of the $U(1)_{\rm X}$, that is generated by a scalar $\sigma$ with charge $Q^{\rm X}_\sigma=-2$ taking a vacuum expectation value $v_\sigma$. Within this setup, inelastic scatterings in the case of scalar DM will generally have a momentum-dependent suppression, and thus the remainder of the discussion will be focused on fermionic DM.

Starting with a Dirac fermion $\Psi$, a small Majorana mass can be generated from the term
\be
\mathcal{L}_{\rm Y} \subset  y_{\rm \Psi}\,\sigma\,\overline{\Psi^c}\,\Psi\,+{\rm h.c.}
\ee
The mass eigenstates are then two Majorana particles $\psi_{1,2}$ with a mass splitting
\be
m_{\psi_2}-m_{\psi_1}\sim y_{\rm \Psi}\,v_\sigma\,.
\ee
We will consider both SI and SD cross sections in what follows. A spin-independent (SI) cross section will be generated by the vector interaction arising from the kinetic term:
\be
\overline\Psi\,\gamma^\mu\,D_\mu\,\Psi \subset - ig \,A'_\mu\,(\psi_1^\dagger\,\bar{\sigma}^\mu \,\psi_2-\psi_2^\dagger\,\bar{\sigma}^\mu \,\psi_1)\,,
\ee
where $\bar\sigma^\mu=(1,-\sigma^i)$ (and $\sigma^\mu=(1,\sigma^i)$).

It is discussed in ref.~\cite{Kopp:2009qt} that the only contribution to an unsuppressed inelastic spin-dependent (SD) cross section comes from the tensor operator $(\overline\Psi\,\Sigma^{\mu \nu}\,\Psi)\,(\overline q\,\Sigma_{\mu \nu}\,q)$, with:
\be
\overline\Psi\,\Sigma^{\mu \nu}\,\Psi \subset -2i \left( \psi^{T}_{2} \sigma^{\mu \nu} \psi_{1} + \psi^{\dagger}_{2} \bar{\sigma}^{\mu \nu} \psi^{*}_{1} \right)\,,
\ee
where $\Sigma^{\mu \nu}=i[\gamma^\mu,\gamma^\nu]/2$, $\sigma^{\mu\nu}=i(\sigma^\mu\bar\sigma^\nu-\sigma^\nu\bar\sigma^\mu)/4$ and $\bar{\sigma}^{\mu\nu}=i(\bar\sigma^\mu\sigma^\nu-\bar\sigma^\nu\sigma^\mu)/4$. 

In the discussion that will follow, it is important to point out that in the general inelastic DM case, there are multiple reasons to expect elastic scattering cross sections. For instance, there can be elastic interactions with SM quarks via mixing of $\sigma$ with the SM Higgs, which will be subdominant with respect to inelastic interactions for $m_{A'} \ll m_\sigma$. Elastic interactions mediated by ${A'}$ are generated at the level of $y_{\rm \Psi}\,v_\sigma/m_{\psi_{1,2}}\sim10^{-6}$ for the fermionic DM case. The inelastic interactions also generate at one loop a subdominant elastic scattering cross section.

Of importance to our discussion is whether the excited state is stable or not. Interactions with the SM occur through the kinetic mixing of $A'$ with the photon and Z-boson~\cite{Pospelov:2007mp}
\be
\mathcal{L} \subset \kappa\, A'_{\mu \nu}\, F^{\mu \nu} + \kappa_Z\, A'_{\mu \nu}\, Z^{\mu \nu}\,.
\ee
We will consider mass-splittings below the MeV range which implies that the possible decays of the excited state will be loop channels into its lower mass counter-part and a pair of neutrinos or into 3 photons, unless there is a very light new particle of mass $<\delta$ into which $\chi^*$ can decay. It is shown in ref.~\cite{Batell:2009vb} that it is quite plausible for the excited state to have a lifetime longer than the age of the Universe for small enough splittings and a heavy mediator such as the ones considered here. The galactic abundance of DM is then made up of both $\chi$ and $\chi^*$ if the relic density of the latter is not depleted by self-scattering processes such as $\chi^*\chi^*\rightarrow \chi\chi$ in the early Universe, i.e., if these processes become inefficient at temperatures above the splitting $\delta$, so that the population of excited states is not Boltzmann suppressed. For heavy enough mediators\footnote{The exact lower bound on the mediator mass depends on the DM mass and on the particular model.} it is thus expected to have both states present today with similar abundances in this type of models.

For the case in which $\chi^*$ is stable, the DD rates are dominated by its contribution via the t-channel reactions $\chi^*\, N\rightarrow \chi \,N$ and the endothermic contribution can be neglected. Thus the contributions of the low mass state (endothermic DM) via the t-channel reactions $\chi\, N\rightarrow \chi^* \,N$ will be relevant when the excited state is unstable, decaying promptly into the lower state and another very low mass particle ($<\delta$) of the dark sector. In these scenarios another important constraint comes from big bang nucleosynthesis. In particular, strong lower bounds on $\kappa$ and $\kappa_Z$ exist depending on the mass of the mediator in order for it to decay early enough. This is crucial for mediators with masses above the MeV, and can yield tensions with DD upper limits.

For the rest of the paper, the endothermic and exothermic cases will be treated independently, keeping in mind that if exothermic scatterings occur, the lower mass state will be populated in an equal or larger amount as the excited state, and thus the whole dark sector contribution to the scattering rate in DD and to the capture rate in the Sun will be given by the sum of both the endothermic and the exothermic components.


\section{Capture rate in the Sun} \label{s:capt}
In this section we will discuss the capture process of inelastic DM in the Sun using the notation of ref.~\cite{Blennow:2015oea}. The case of elastic capture was originally considered in refs.~\cite{Press:1985ug, Griest:1987yu, Gould:1987ir} (see also refs.~\cite{Ullio:2000bv, Peter:2009mk, Serpico:2010ae, Liang:2013dsa,Choi:2013eda,Kavanagh:2014rya,Arina:2013jya,Kappl:2011kz,Wikstrom:2009kw,Hooper:2008cf,Kamionkowski:1994dp,Bergstrom:1998xh,Ibarra:2014vya,Catena:2015iea} for other more recent studies) and expanded to cover inelastic dark matter in refs.~\cite{Nussinov:2009ft, Menon:2009qj, Shu:2010ta}.

\subsection{Kinematics}
When a DM particle falls into the gravitational potential of the Sun, it will accelerate to a velocity $w=\sqrt{v^2+u_{\rm esc}^2}$, where $v$ is the velocity of the DM particle outside the potential and $u_{\rm esc}$ is the local escape velocity of the Sun. For endothermic scattering of DM particles to be possible, the center of mass energy of the DM--target system must be larger than $\delta$, leading to the constraint:
\be
v^2 > v^2_{\rm lower} = 2\delta/ \mu - u_{\rm esc}^2\,,
\ee
where $\mu$ is the reduced mass of the DM--target system. This constraint has a large impact on the solar capture. For $\delta \sim \mathcal{O}(100)$~keV, $v$ must be extremely large for the lighter elements to possibly capture the DM particles. For this reason, capture on hydrogen does not occur at all unless $\delta$ is very small. This, together with the very low solar abundance of heavy isotopes with spin, results in a negligible SD capture, see~\cite{Shu:2010ta}. On the contrary, for exothermic DM, there is no constraint on the incoming velocity which implies that capture is possible also through SD interactions. We will henceforth consider only hydrogen for capture through SD cross sections.

In an inelastic collision, the maximum and minimum recoil energy of the target nucleus depend on the incoming DM particle velocity $v$ as
\be \label{Emax_in}
E_{\rm max}(w)=\frac{\mu^2}{m_{\rm A}} w^2 \left( 1+\sqrt{1-\frac{\delta}{\mu w^2/2}} \right) -\frac{\mu}{m_{\rm A}} \delta\,,
\ee
\be \label{Emin_in}
E_{\rm min}(w)=\frac{\mu^2}{m_{\rm A}} w^2 \left( 1-\sqrt{1-\frac{\delta}{\mu w^2/2}} \right) -\frac{\mu}{m_{\rm A}} \delta\,.
\ee
The minimum velocity $v_{\rm m}$ that can result in a recoil energy $E_{\rm R}$ is obtained by solving the equations~\eqref{Emax_in}~and~\eqref{Emin_in} for $v$,
\be \label{v_m}
v^2_{\rm m} = \left( \sqrt{\frac{m_{\rm A}E_{\rm R}}{2\mu^2}} + \frac{\delta}{\sqrt{2 m_{\rm A} E_{\rm R}}} \right)^2 - u_{\rm esc}^2\,.
\ee
The recoil energy of the target nuclei after the DM scattering is uniformly distributed over the energy interval $[E_{\rm min}(w), E_{\rm max}(w)]$. In order to be captured, the interaction must leave the scattered DM particle with a velocity less than the local escape velocity, which in turn implies that the recoil energy of the nucleus must be larger than
\be \label{Emincapt}
E_{\rm capt}(v) = m_\chi v^2/2 - \delta\,.
\ee
Thus, for a DM particle to be captured, its velocity has to be larger than $v^2_{\rm capt}=2 \delta/m_\chi$, which corresponds to zero recoil energy. 

The inelastic cross section is accompanied by a phase-space factor and can be written as $\sigma_{{\rm inel}}$ = $\sqrt{1-2 \delta/\mu w^2}\, \sigma_{{\rm elast}}$. We take $\kappa = f_n / f_p$ as the ratio of the DM--neutron coupling and the DM--proton couplings and use that $\sigma_{\chi \rm A} \propto A^2_{\rm eff}\, \sigma_{\chi \rm p}$, with $A_{\rm eff} = Z + \kappa \,(A-Z)$, where we have absorbed the proton coupling in the cross section. The differential cross section can now be expressed as
\begin{align} \label{differential_CS}
\frac{d\sigma_{\rm A}}{dE_{\rm A}}(w,E_{\rm R}) &= \sqrt{1-\frac{\delta}{\mu v^2/2}}\, \frac{1}{E_{\rm max}(w)-E_{\rm min}(w)}\, A_{\rm eff}^2 \, \frac{\mu^2}{\mu_{\chi p}^2} \, \sigma_{\chi p} \, F^2_{\rm A}(E_{\rm R}) \nonumber \\ &= \frac{m_{\rm A} A_{\rm eff}^2 \sigma_{\chi p}}{2 \mu_{\chi p}^2 w^2} F^2_{\rm A}(E_{\rm R})\,,
\end{align}
where the phase space factor was cancelled by the denominator and we are left with the same differential cross section as in the elastic case.

\begin{figure}
	\centering
	\includegraphics[width=0.5\textwidth]{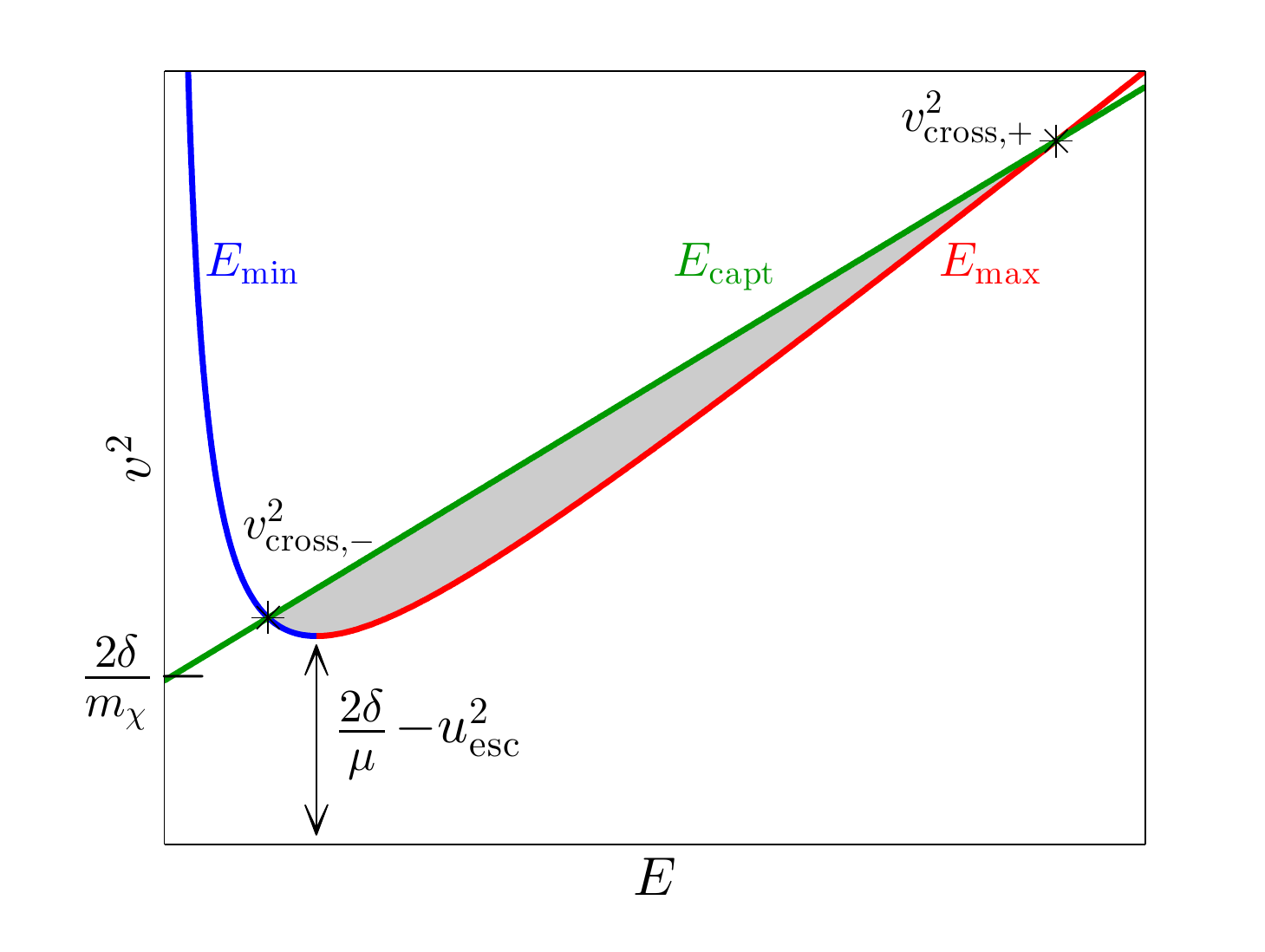}~~
	\caption{$v^2$ as a function of $E_{\rm min}$ (blue line), $E_{\rm max}$(red line) and $E_{\rm capt}$ (green line) for the general endothermic case. In this case $v^2_{\rm lower} = 2\delta/ \mu - u_{\rm esc}^2>v^2_{\rm capt}=2 \delta/m_\chi$. The shaded area represent the one over which the horizontal integration to obtain $C_{\rm Sun}$ is performed.} \label{v2_Er}
\end{figure}

A key input for computing the DM scattering rate is the velocity distribution. We will denote this distribution in the solar frame by $f(\vec{v})$ and normalize it as
\be \label{veldist}
\int \, d^3v \, f(\vec{v}) = \int_0^\infty \, dv \, v^2 \, \tilde{f}(v) = 1\,,
\ee
where we have defined the angular averaged velocity distribution as $\tilde{f}(v) = \int  d\Omega \, f(v,\Omega)$, with $d\Omega = \sin\theta\, d\theta\, d\phi$. The capture rate of DM particles in the Sun in the inelastic case is given by~\cite{Gould:1987ir}
\beq \label{capture}
C_{\rm Sun}= 4\pi \, \mathcal{C}\sum_{A} A_{\rm eff}^2 \int_0^{R_{\rm Sun}} dr r^2 \rho_A(r) \int_{v_{\rm min}^A (r)}^{v^A_{\rm max} (r)} dv \tilde{f}(v)\,v\, \mathcal{F}_A(v,r)\,,
\eeq
where the sum runs over all elements present in the sun and $\rho_A(r)$ is their number density at a specific radius.  We have also defined
\be
\mathcal{F}(v,r) = \int_{E_{\rm m}(v)}^{E_{\rm max}(v)} F_{\rm A}^2 (E_{\rm R})\, dE_{\rm R}\,,
\ee
where $E_{\rm max}(v)$ is the maximum possible recoil in a collision and $E_{\rm m}(v)$ as the smallest recoil energy resulting in the capture of the DM particle.
The function $F^2_A(E_{\rm R})$ is a nuclear form factor for which we use the approximation $F^2_A(E_{\rm R}) \sim e^{-E_{R}/E_A}$ with $E_A = 3/2 m_A R_A^2$ and $R_A = \left[ 0.91(m_A/{\rm GeV})^{1/3} + 0.3 \right]$ fm~\cite{Gould:1987ir}. The constant $\mathcal{C}$ is a combination of the DM--proton scattering cross section $\sigma_{\chi p}$, the reduced mass $\mu_{\chi p}$ and the local DM energy density $\rho_\chi$, given by
\be \label{C}
\mathcal{C} = \frac{\rho_\chi \sigma_{\chi p}}{2 m_\chi \mu^2_{\chi p}}\,.
\ee
Finally, it is also necessary to identify the largest velocity for which a DM particle can be captured, $v^A_{\rm max}$, and the smallest one, $v^A_{\rm min}$. Let us start with endothermic scattering. In the special case of  $v_{\rm lower}^2 >v^2_{\rm capt}$, any DM particle that scatters, regardless of the recoil energy, will be captured. Fig.~\ref{v2_Er} shows $v^2$ as a function of $E_{\rm min}$, $E_{\rm max}$ and $E_{\rm capt}$ (eqs.~\eqref{Emax_in}, ~\eqref{Emin_in} and~\eqref{capture} respectively), as well as the region over which one integrates horizontally in eq.~\eqref{capture}.  When there is no cross between $E_{\rm capt}$ and $E_{\rm max}$ or $E_{\rm min}$, the recoils are too weak to capture the DM particle. When $E_{\rm capt}$ cross $E_{\rm max}$ twice, there will be one region in velocity to integrate over given by $\left[ v^A_{\rm min} = v_{\rm cross,-},v^A_{\rm max}=v_{\rm cross,+} \right]$ in which case $E_{\rm m} = E_{\rm capt}$. When $E_{\rm capt}$ crosses $E_{\rm min}$ and $E_{\rm max}$ once each, there will be one additional region to integrate over on top of the one previously mentioned, given by $\left[ v^A_{\rm min} = v_{\rm lower},v^A_{\rm max}=v_{\rm cross,-} \right]$, where $E_{\rm m} = E_{\rm min}$. For $v_{\rm lower}^2 <v^2_{\rm capt}$, one can see in the figure that there will always be the two mentioned regions to integrate over. Finally, for $v_{\rm lower}^2 < 0$ the lower integration limit is of course $v^A_{\rm min}=0$. For exothermic DM the discussion is identical except that $v^A_{\rm min}=0$ always applies.

\begin{figure}
	\centering
	\includegraphics[width=0.5\textwidth]{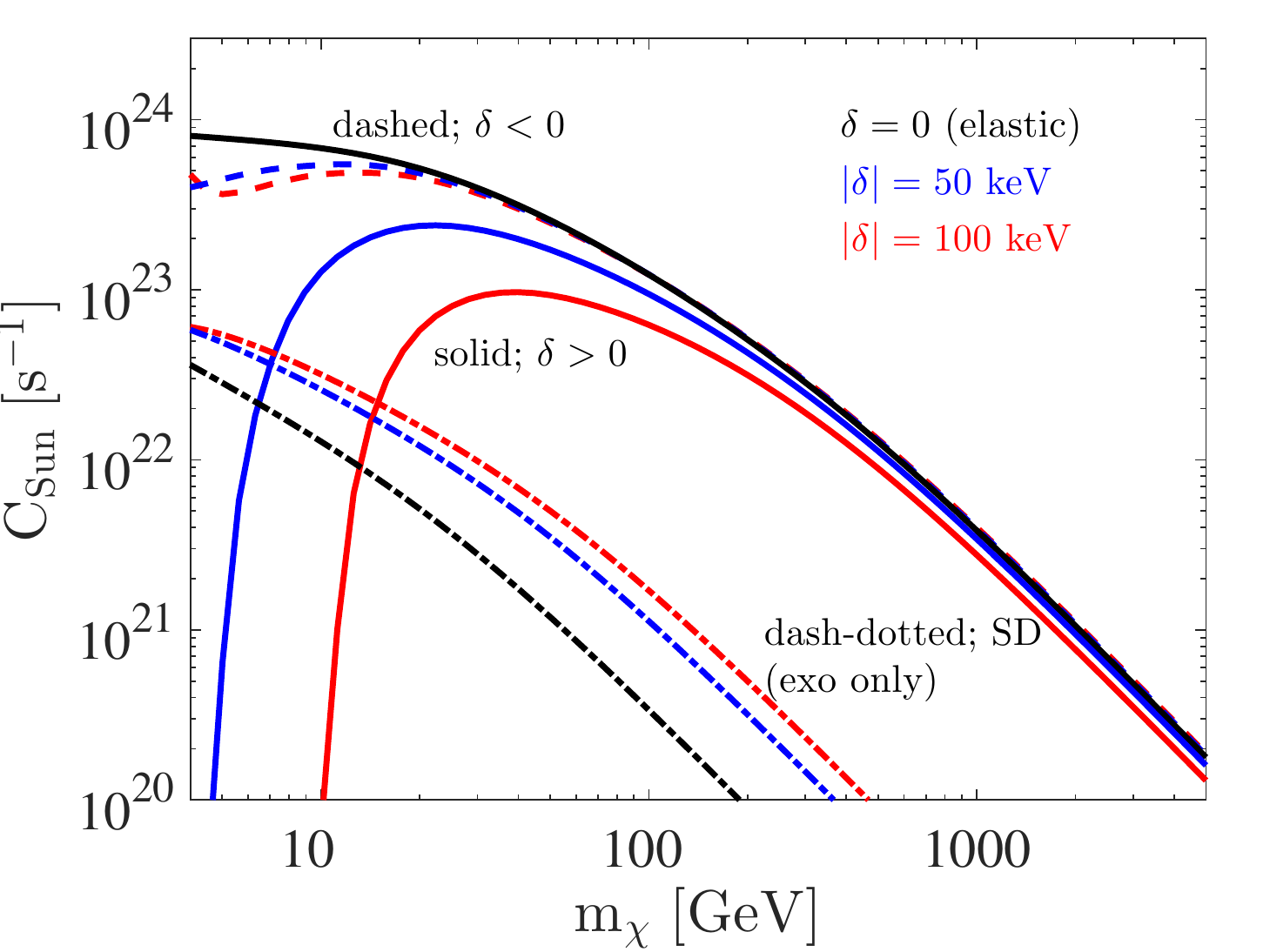}~~
	\includegraphics[width=0.5\textwidth]{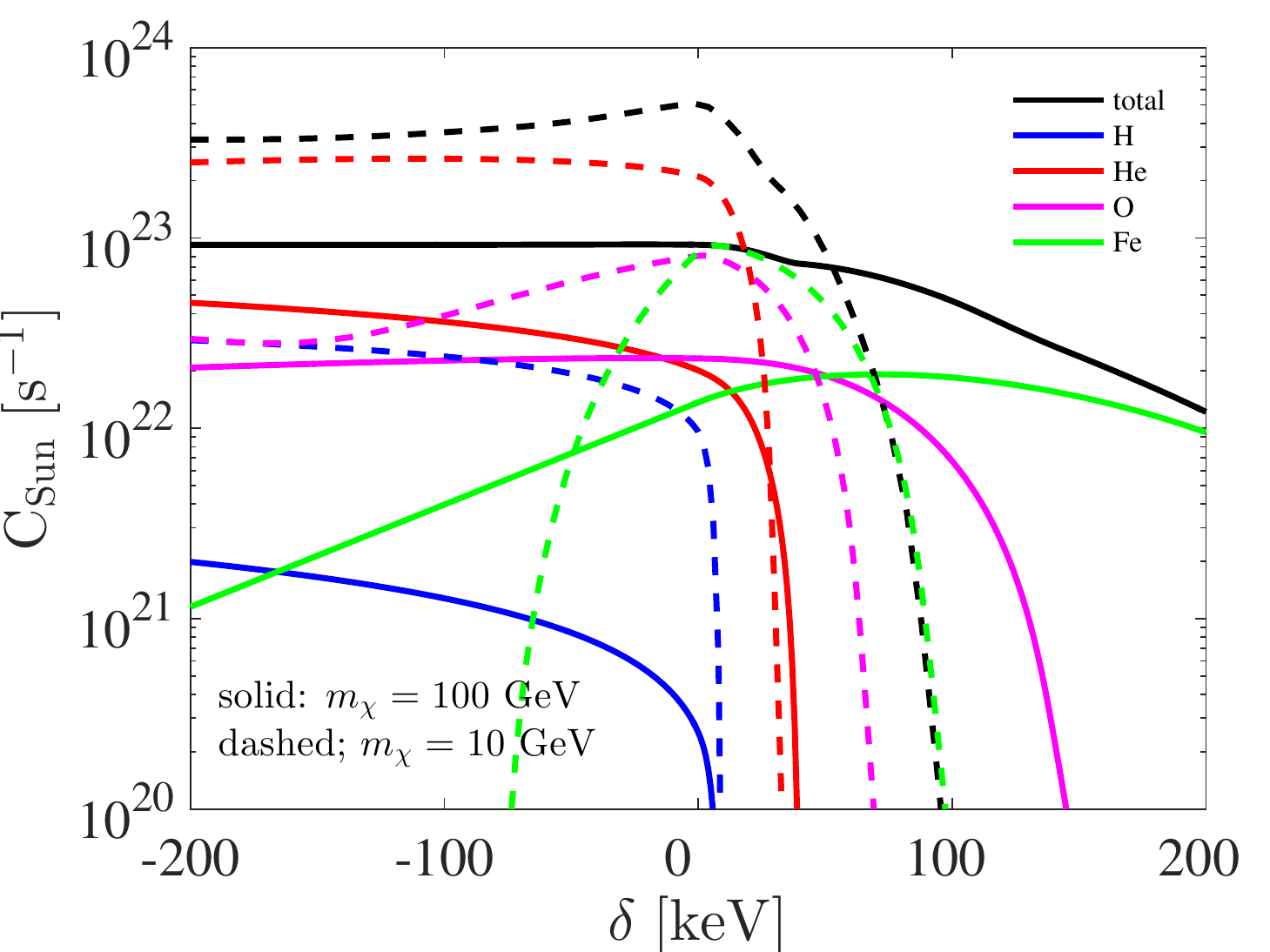}
	\caption{Left) We show the DM capture rate as a function of mass for $\delta=50$~keV (blue) and $\delta=100$~keV (red) for $\sigma=10^{-42}$~cm$^2$. We show as solid lines the endothermic case and in dashed (dash-dotted) the SI (SD) exothermic cases. Right) Total capture rate as a function of $\delta$ as well as the contribution from various elements for $m_{\chi}=10\, (100)$~GeV as dashed (solid) lines.} \label{caprates}
\end{figure}

The velocity $v_{\rm cross,\pm}$ is found by setting $E_{\rm max}$ and $E_{\rm min}$ equal to $E_{\rm capt}$ and solving for $v$, yielding the rather uninformative solutions:
\begin{align} \label{vcross}
v_{\rm cross,\pm}^A &= \frac{\sqrt{2\,m_\chi \,m_A}}{|m_\chi-m_A|}\,u_{\rm esc}(r)\, \sqrt{1-\delta \frac{m_\chi-m_A}{m_\chi^2\,u^2_{\rm esc}(r)}      \pm      \sqrt{1- 2 \delta \frac{m_\chi-m_A}{m_\chi m_A\,u^2_{\rm esc}(r)}}}\,.
\end{align}
The capture rates are shown in fig.~\ref{caprates} for the conventional SHM\footnote{In this paper we will use the solar speed $v_{\rm Sun}=220$~km/s, the galactic escape velocity $v_{\rm esc}=544$~km/s, and the local DM density $\rho_{\chi}=0.4$~GeV/cm$^3$~\cite{Catena:2009mf,Weber:2009pt,Salucci:2010qr,Pato:2010yq,Iocco:2011jz,Garbari:2012ff,Bovy:2012tw}.} as a function of DM mass ($\delta$) in the left (right) plot. For large DM masses, the capture rates with a SI cross section are very similar, while for lower masses capture of endothermic DM is very inefficient due to the center of mass energy not being large enough to excite the DM state. In the right plot one can see that in the endothermic case, the heavy elements (in particular iron but oxygen is also important) dominate the capture. For the low DM masses, helium may overtake iron as the main captor for small $\delta$. In the exothermic case, helium is the strongest captor.

\subsection{Thermalization}

In order to compute the annihilation rate of DM in the Sun, it is necessary to know or make reasonable assumptions on the captured DM distribution. This distribution is usually taken as an isothermal Maxwell--Boltzmann distribution with a temperature given by the temperature at the solar center, $n(r) \sim e^{-E/ (k_B\,T_c)}$. The validity of this approximation rests on the assumption that the DM particles scatter repeatedly in the Sun and eventually reach thermal equilibrium with the solar matter. An important aspect of the capture process is therefore not only the initial capturing interaction, which will generally leave the captured particle with an energy larger than the thermal energy, but also how the DM continues to scatter and eventually thermalizes. In our case, the thermal distribution will mainly contain the lower mass state as long as $\delta > k_B\,T_c$, since the population of the higher mass state will be thermally suppressed.\footnote{The elastic annihilations can thus be comparable to the inelastic ones.}

Regardless of whether the capture is endo- or exothermic, subsequent scatters will continuously change the state of the captured particle between $\chi$ and $\chi^*$. In this process, there is an increased probability of losing the captured particle due to the energy release when the higher mass state interacts and creates a particle of the lower mass state. In the case of endothermic dark matter, this can be partially avoided if the $\chi^*$ decays sufficiently fast while, for exothermic dark matter, the first such interaction is the second interaction after capture and the DM particle has already lost additional energy in the first post-capture interaction, which decreases the risk of ejecting the particle. As such, we do not expect this ejection to change the actual capture rate significantly.

Once the DM particle has lost enough energy, the up-scatter process to the higher mass state will no longer be kinematically available. From that point on, the lower mass state can only lose energy through a residual elastic process which is suppressed with respect to the inelastic one, see, e.g., the discussion in sec.~\ref{s:pheno} (and for instance ref.~\cite{Cui:2009xq}). A simple estimate of the elastic cross section required for thermalisation on protons is~\cite{Menon:2009qj}
\be
\sigma_{\rm elast} \gtrsim 10^{-50} \left( m_\chi/{\rm GeV} \right) \, \rm cm^2.
\ee
Note that this is a conservative estimate for two reasons. DM will initially lose energy through inelastic scattering and the average energy loss due to scattering on heavier elements is greatly increased. Due to the cross sections we will consider, this leaves a large range for an elastic cross section sizeable enough for thermalisation by elastic scattering while having negligible contribution to the capture rates, as can be seen in fig.~\ref{caprates}. We assume, along the lines of refs.~\cite{Nussinov:2009ft,Menon:2009qj}, that there is a sizeable enough elastic cross section for thermalisation to occur.

\subsection{Equilibrium and the neutrino signal}
The number of captured DM particles in the Sun, $N$, is governed by the differential equation
\be
\dot{N} = C_{\rm Sun} - C_{\rm ev}N - C_{\rm ann}N^2\,,
\ee
where $C_{\rm ev}\,N$ is the evaporation rate and the last term is equal to twice the annihilation rate, i.e., $C_{\rm ann}\,N^2=2\,\Gamma_{\rm Sun}$. We will conservatively consider DM masses of $m_\chi \geq 10$~GeV, in which case evaporation has been shown to be negligible~\cite{Griest:1987yu,Gould:1987ju,Busoni:2013kaa}. The annihilation rate is then given by
\be
\Gamma_{\rm Sun}  = \frac{1}{2}\, C_{\rm Sun} {\rm tanh} \left( \frac{t}{\tau} \right)\,,
\ee
with $\tau = 1/\sqrt{C_{\rm Sun} \,C_{\rm ann}}$. For $t > \tau$, equilibrium has been reached and the annihilation rate is given by
\be \label{equilibrium}
\Gamma_{\rm Sun} = \frac{1}{2} C_{\rm Sun}\,.
\ee
Equilibrium requires that $\tau > t_{\odot}$ which is related to the capture rate as (see for instance ref.~\cite{Nussinov:2009ft}):
\be \label{equilibrium2}
\tau \lesssim 0.1 \, t_\odot \left( \frac{10^{21} \, {\rm s^{-1}}}{C_{\rm Sun}} \right)^{1/2} \left( \frac{3 \cdot 10^{-26} \, {\rm cm^3\,s^{-1}}}{\langle \sigma v \rangle} \right)^{1/2} \left( \frac{100 {\rm GeV}}{m_\chi} \right)^{3/4}\,,
\ee
where $\langle \sigma v \rangle$ is the thermally averaged cross section. It is thus important to remember that equilibrium has only occurred for capture rates larger than $\sim 10^{21}$~s$^{-1}$, or for an annihilation cross section that is enhanced with respect to the thermal freeze-out one. If the Sun effectively captures DM, its subsequent annihilation into SM particles will produce neutrinos at a rate that may be detectable on neutrino telescopes. The flux produced by annihilations into particles $f$ with branching ratio ${\rm BR}_f$ is given by
\be
\frac{d\phi^f_\nu}{dE_\nu} = {\rm BR}_f \frac{\Gamma_{\rm Sun}}{4 \pi d^2} \frac{d N^f_\nu}{dE_\nu}\,,
\ee
where $d$ is the distance between the Sun and the Earth and $d N^f_\nu/dE_\nu$ is the neutrino spectrum for flavour $f$ to reach the earth, taking into account neutrino oscillations, the neutrino-matter effects and so on~\cite{Blennow:2007tw,Cirelli:2005gh}.

\section{Direct detection} \label{s:DD}

In this section we review direct detection~\cite{Goodman:1984dc} of inelastic DM~\cite{TuckerSmith:2001hy,TuckerSmith:2004jv}. The differential rate can be written as:\footnote{In the following we assume that the detector is made-up of one element, or in the case of multi-target detectors, that one element dominates the rate.}
\be
\mathcal{R}(E_{\rm R},t) =\frac{\rho_\chi}{m_\chi m_{A}} \int_{| \vec{v} | > v_m} d^3v \, v f_{\rm det}(\vec{v}) \,\frac{d\sigma_{A}}{dE_{\rm R}}\,,
\ee
where $v_{\rm m}$ is the lowest velocity that can produce a recoil $E_{\rm R}$ in the detector,  given by eq.~\eqref{v_m} without the $u_{\rm esc}^2$ term.
It is useful to define $\tilde{\eta}(v)$ as:
\beq \label{eq:eta} 
\tilde{\eta}(v_{\rm m}, t) \equiv \mathcal{C}\,\eta(v_{\rm m}, t) \quad\text{with}\quad \eta(v_{\rm m}, t) \equiv 
\int_{v_{\rm m}}^\infty  d v \,v \tilde{f}_{\rm det} (v,t)\,,
\ee
where $\mathcal{C}$ is given by eq.~\eqref{C}.

The velocity distribution in the detector rest frame is related to the velocity distribution in the solar frame $f(\vec{v},t)$  by a simple Galilean transformation, $f_{\rm det}(\vec{v}) = f(\vec{v} + \vec{v}_{e}(t))$, where $\vec{v}_{e}(t)$ is the velocity of the earth in the solar frame, which is expected to give rise to a modulation in the rate seen by an experiment~\cite{Drukier:1986tm, Freese:1987wu} (see refs.~\cite{HerreroGarcia:2011aa, HerreroGarcia:2012fu, Herrero-Garcia:2015kga} for halo-independent bounds for annual modulation signals). In the following we will neglect the small dependence of the velocity distribution on the velocity of the Earth and assume that the solar and detector distributions are the same, i.e., $f_{\rm det}(\vec{v})=f(\vec{v})$. Using the differential cross section of eq.~\eqref{differential_CS}, now as a function of $v$, the differential event rate takes the form
\be
\mathcal{R}(E_{\rm R}) =  A^2 F^2_A(E_{\rm R}) \tilde{\eta}(v_{\rm m},t)\,.
\ee
The number of events in a detector in an energy range $[E_1,E_2]$ is given by
\be
N_{\rm [E_1,E_2]} = MTA^2 \int_{0}^{\infty} dE_{\rm R} F_A^2(E_{\rm R}) G_{\rm [E_1,E_2]}(E_{\rm R}) \tilde{\eta}(v_{\rm m},t)\,,
\ee
where $M$ is the detector mass, $T$ is the exposure time, $G_{\rm [E_1,E_2]}(E_{\rm R})$ is the detector response function describing the probability that a DM event with true recoil energy $E_{\rm R}$ is reconstructed in the energy interval $\rm [E_1,E_2]$ including energy resolution, energy dependent efficiencies, and possibly also quenching factors.

Once a differential rate $\mathcal{R}(E_{\rm R})$ is detected, one has direct information on $\tilde{\eta}(E_{\rm R})$. For a specific DM mass, $\mathcal{R}(E_{\rm R})$ can be translated into $\mathcal{R}(v_{\rm m})$ using eq.~\eqref{v_m} (without the $u_{\rm esc}^2$ term). The extracted $\tilde{\eta}(v_{\rm m})$ must then be the same in all experiments, which is the key observation of halo-independent methods~\cite{Fox:2011bu,Fox:2011bz}.\footnote{See also refs.~\cite{Bozorgnia:2013hsa, McCabe:2011sr, McCabe:2010zh, Frandsen:2011gi, HerreroGarcia:2011aa, HerreroGarcia:2012fu, DelNobile:2013cta, DelNobile:2013cva, Bozorgnia:2014gsa, Fox:2014kua, Feldstein:2014ufa, Cherry:2014wia} for examples of the use of this method to compare results in different DD experiments and refs.~\cite{Blennow:2015gta, Herrero-Garcia:2015kga, Ferrer:2015bta} for other methods to combine DD signals with other searches independently of the astrophysics.} 

We will make use of the fact that, once a differential rate is observed and $\tilde{\eta}(v_{\rm m})$ is extracted, information of the velocity distribution above $v_m$ can be obtained from $\tilde{\eta}$ using the relation~\cite{Drees:2007hr}
\be \label{extr_veldist}
\mathcal{C}\tilde{f}_{\rm extr}(v) = -\frac{1}{v}\frac{d\tilde{\eta}(v)}{dv} = -\frac{1}{vA^2_{\rm eff}} \frac{d}{dv} \left( \frac{R(E_{\rm R})}{F^2_A(E_{\rm R})} \right)\,,
\ee
and therefore a lower bound on the capture rate can be obtained~\cite{Blennow:2015oea}.

\begin{figure}
	\centering
	\includegraphics[width=0.5\textwidth]{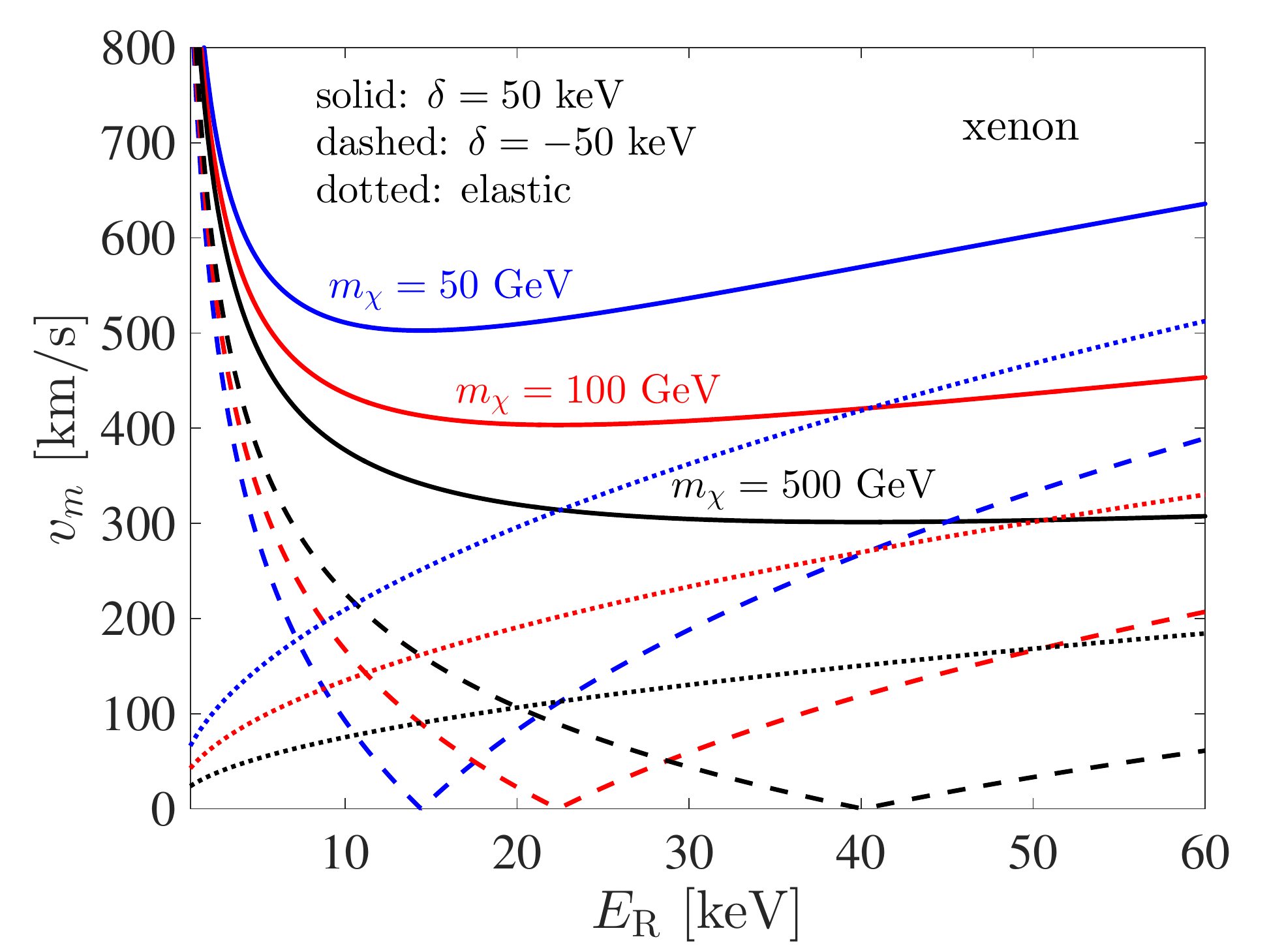}~~
	\includegraphics[width=0.5\textwidth]{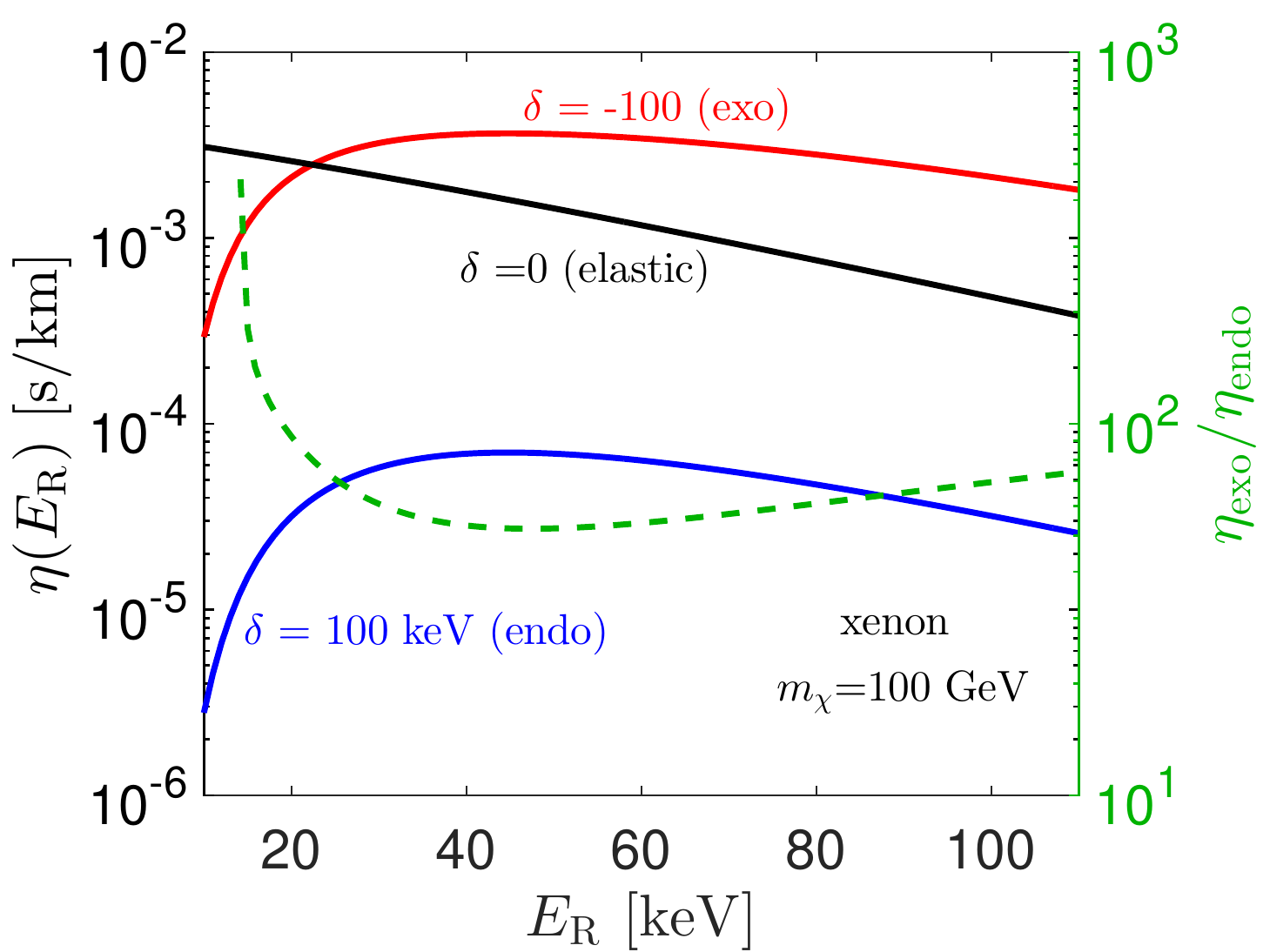}
	\caption{Left) We show $v_m$ versus the recoil energy for xenon in the case of endothermic (solid) and exothermic (dashed) interactions with $\delta=50\, (-50)$ keV, respectively. We show results for $m_\chi= 50,100,500$ GeV in blue, red and black respectively (from top to bottom). Right) $\eta(E_{\rm R})$ versus recoil energy for endothermic, exothermic and elastic interactions for a DM mass $m_{\chi}=$~100~GeV and $|\delta |=100$~keV. The ratio $\eta_{\rm exo}/\eta_{\rm endo}$ is shown in green dashed.} \label{vminxe}
\end{figure}

In the case of elastic scattering, one can probe the velocity distribution down to a velocity $v_{\rm thr}$ set by the threshold energy of the experiment, $E_{\rm thr}$ (eq.~\eqref{v_m} without the $u_{\rm esc}^2$ term and $\delta=0$). If DM scattering is inelastic, the relationship between $E_{\rm R}$ and $v_{\rm m}$ is no longer unique, see for instance ref.~\cite{Bozorgnia:2013hsa}. Depending on the values of $m_{\chi}$ and $\delta$, certain values of $v_{\rm m}$ correspond to two values of $E_{\rm min}$, while others correspond to none. There is however a minimum velocity $(v_{\rm m})_{\rm min}$ for which recoils occur, given by
\beq \label{minvmin}
(v_{\rm m})_{\rm min}=\sqrt{2|\delta|/\mu_{\chi A}}\,,\qquad\text{at the energy }\qquad E_{\rm min} = \mu_{\chi A}|\delta|/m_A\,.
\eeq
The left panel of fig.~\ref{vminxe} shows $v_{\rm m} (E_R)$ for a xenon target versus the recoil energy for various DM masses, with a mass splitting $\delta=50$~keV. In the endothermic case, any particle with a velocity less than $(v_{\rm m})_{\rm min}$ cannot be detected. This in turn implies that the differential rate works as a probe only in the velocity region $v > (v_{\rm m})_{\rm min}$. In exothermic models, $v_{\rm m}=0$ at some particular energy $E^{\rm obs}_{\min}$, and thus the full velocity distribution will be probed by the experiment as long as the threshold energy is low enough, c.f. fig.~\ref{vminxe}.

For inelastic interactions a maximum in the spectrum of $\tilde{\eta}(E_{\rm R})$ is expected at some particular energy. From the observed spectrum we can derive the minimum energy $E^{\rm obs}_{\min}$ where $\tilde{\eta}(E_{\rm R})$ is maximal and from eq.~\eqref{minvmin} obtain $\delta$ as a function of the DM mass:
\be \label{dm}
|\delta\,(m_{\chi})| = \frac{m_{\rm A}}{\mu} E^{\rm obs}_{\min}\,,
\ee
which is thus uniquely defined, with only the DM mass being a free parameter. In the case that a signal is found in an experiment and can be explained by inelastic DM scattering, eq.~\eqref{dm} implies that $\delta$ will always larger than $E^{\rm obs}_{\rm min}$. 

In the right panel of figure~\ref{vminxe} we plot $\eta(E_{\rm R})$ versus $E_{\rm R}$  for elastic (black), endothermic for $\delta=100$~keV (blue) and exothermic for $\delta=-100$~keV (red) for $m_{\chi}=100$~GeV. As can be seen in fig.~\ref{vminxe}, the expected maximum in the inelastic spectrum lies at the same recoil energy for both endo- and exothermic. Notice that the probability to measure the upturn increases greatly with the energy range in which the rate is measured. 

It should be noted that $E^{\rm obs}_{\rm min}$ does not correspond to the recoil energy at which the rate $\mathcal{R}(E_R)$ reaches its maximum. The energy dependence of the form factor or the efficiencies/resolutions imply that the maximum of the event rate and of $\eta (E_{\rm R})$ will be at different energies. Uncertainties in these quantities thus propagate to $E^{\rm obs}_{\rm min}$ and will affect the extraction of the DM parameters such as the mass or the splitting, and thus a precise knowledge of them is of uttermost importance. In order to illustrate our strategy, in the following we will deal with an idealized situation and assume that one can extract the precise energy value at which $\tilde{\eta}(E_R)$ is maximum, exactly unfolding the form factors, energy resolutions, efficiencies, quenching factors and other experimental features. Of course, once a real signal is observed, a proper statistical treatment would have to be performed.

In the plot of figure~\ref{vminxe} it is also shown for illustrative purposes the ratio of $\eta_{\rm exo}(E_R)$ and $\eta_{\rm inel}(E_R)$ as dashed green. One can see that the ratio between $\eta_{\rm exo}(E_R)$ and $\eta_{\rm inel}(E_R)$ takes values larger than $\sim 30$. The diverging behaviour for low recoil energies is due to the exponential suppression for the velocities required for endothermic scattering, which are closer to the escape velocity than those of exothermic scattering. Notice that assuming the SHM, by observing the low energy behaviour, in principle one may be able identify the inelastic or exothermic nature of an observed spectrum.

An important point is that larger rates are expected for the exothermic case. For the fixed values of the DM parameters used in this example, we have $\eta_{\rm exo}\gg \eta_{\rm endo}$, and thus the expected number of events for exothermic DM can be roughly $\sim1-2$ orders of magnitude larger than for endothermic scattering. This implies that if both states are present, exothermic scattering will typically provide the dominant contribution to the rates. 

Once a DD signal in a direct detection experiment is observed, one should first check if the rate is compatible with inelastic scattering. In ref.~\cite{Bozorgnia:2013hsa} halo-independent methods were devised for that goal, and in particular \emph{the shape test} allows one to test if a DD signal is compatible with inelastic DM. This uses the fact that the $\tilde{\eta}(v_{\rm m})$ extracted from the rate of the two energy branches corresponding to the same $v_{\rm m}$ should coincide. In the following we will assume that a DD signal is detected and that the spectrum fulfils the {\it the shape test} at some confidence level.

\subsection{Combining two positive direct detection signals} \label{2signals}

In this subsection we want to discuss and emphasize the importance of measuring a signal in two independent experiments with different target nuclei. We assume that both are compatible with \emph{the shape test}. Then one will find two different recoil energies in the two experiments for which $\tilde{\eta}(E_R)$ is maximal. Since the observed energy from an spectrum will correspond to a specific $\delta$ in terms of the DM mass according to eq.~\eqref{dm}, one can, from two independent signals, find both the DM mass and the absolute value of the splitting delta. If an energy $E_1^{\rm obs}$ is measured in the first experiment with target nuclei of mass $m_1$, and $E_2^{\rm obs}$ is observed in a second experiment with target nuclei of mass $m_2$, then $m_\chi$ and $\delta$ will be given in terms of the observed energies and target masses by
\be
m_\chi = \frac{m_1 E_1^{\rm obs} - m_2 E_2^{\rm obs}}{E_2^{\rm obs} - E_1^{\rm obs}}, \qquad |\delta| = \frac{E_1^{\rm obs} E_2^{\rm obs}(m_1 - m_2)}{m_1 E_1^{\rm obs} - m_2 E_2^{\rm obs}}\,.
\ee
Notice that for the case in which $m_1$ is larger than $m_2$, $E^{\rm obs}_2$ is necessarily smaller than $E^{\rm obs}_1$, and thus $m_\chi,\, |\delta|$ will be positive. This is a qualitative change with respect to elastic interactions, in which finding the DM mass halo-independently is not so straightforward, even with 2 different signals, see for instance ref.~\cite{Drees:2008bv,Kavanagh:2013wba}. This method however cannot be used for elastic DM, since the maximum of $\tilde{\eta}(E_R)$ is at $E_R = 0$.  

Once the splitting $|\delta|$ and the DM mass $m_\chi$ have been obtained, $\tilde{\eta}$ can be plotted as a function of the minimum velocity $v_{\rm m}$, both for exo- and endothermic interactions, by making use of eq.~\eqref{v_m} (without the $u_{\rm esc}^2$ term) to translate from recoil energies to velocity space. In $v_{\rm m}$ space, if compared in the same velocity range the $\tilde{\eta}$ extracted from both signals should be the same~\cite{Fox:2011bu,Fox:2011bz}. Thus, in principle, one could quantify whether the agreement is better for the exo- or for the endothermic cases, by performing a fit to the spectral shape of $\tilde{\eta}(v_{\rm m})$, which is different due to the different energy dependence of the velocity in both cases, see the left panel of figure~\ref{vminxe}. Thus one could theoretically break the exothermic-endothermic degeneracy. In the following sections we will treat both cases separately.


\section{Numerical analysis} \label{s:bound}

\subsection{Mock data for direct detection}
Under the assumption that a DM DD signal has been observed, a lower bound on the capture rate in the standard elastic DM scenario was derived in ref.~\cite{Blennow:2015oea}. In this section we will extend it to the case of inelastic DM interactions. After having extracted the velocity distribution $\tilde{f}_{\rm extr}(v)$ for velocities above some velocity $v_{\rm thr}$ from a DD signal, the lower bound on the capture rate will be given by
\beq \label{bound}
C_{\rm Sun} \geq 4\pi \, \sum_{A} A_{\rm eff}^2 \int_0^{R_{\rm Sun}} dr r^2 \rho_A(r) \int_{v_{\rm thr}}^{v_{\rm cross} (r)} dv \left( -\frac{d\tilde{\eta}(v)}{dv} \right) \, \mathcal{F}_A(v,r)\,.
\eeq
Of course, if $v_{\rm thr} > v_{\rm cross}$, no lower bound can be set on the capture rate as there is no knowledge of the velocity distribution for those velocities that may result in capture.

To provide examples of the use of the method to find lower bounds on the capture rate, we will generate mock data assuming the conventional SHM for a xenon experiment experiment with natural abundances of isotopes~\cite{Malling:2011va, Baudis:2012bc, Aprile:2012zx,Aprile:2015uzo}. We take the composition of the Sun as reported in ref.~\cite{Serenelli:2009yc}. We take the threshold energy of the xenon experiment at $E_{\rm thr}=3$~keV, the SI cross section to be $\sigma_{\rm SI}=10^{-45}$~cm$^2$ and the SD cross section $\sigma_{\rm SD}=2\cdot 10^{-40}$~cm$^2$. For a DM particle with $m_\chi=100$~GeV and for an exposure of 1 kton yr at 100\% efficiency, about 197 (315) events with recoil energies in the range $3-45$~keV would be observed for elastic SI (SD) interactions. With the same parameters and $\delta=50\,(100)$~keV, an endothermic DM model would imply the observation of 34 (2) events and an exothermic model of $195\,(95)$ events. The upper bound on the endothermic SI cross section is thus relaxed by roughly one and two orders of magnitude respectively for these choices of $\delta$. In the case of SD interactions, the expected number of events is, for $\delta=50\,(100)$~keV: 53 (3) events for endothermic scatterings and 309 (153) events for exothermic scatterings. Notice that for a DM particle with $m_\chi=10\,(100)$~GeV, for $\delta=50$~keV, the energy at which the $\eta(E_{\rm R})$ is maximum, given by eq.~\eqref{dm}, will be $3.7\, (22)$ keV. For $\delta=100$~keV these energies are a factor of 2 larger.

\begin{figure}
	\centering
	\includegraphics[width=0.5\textwidth]{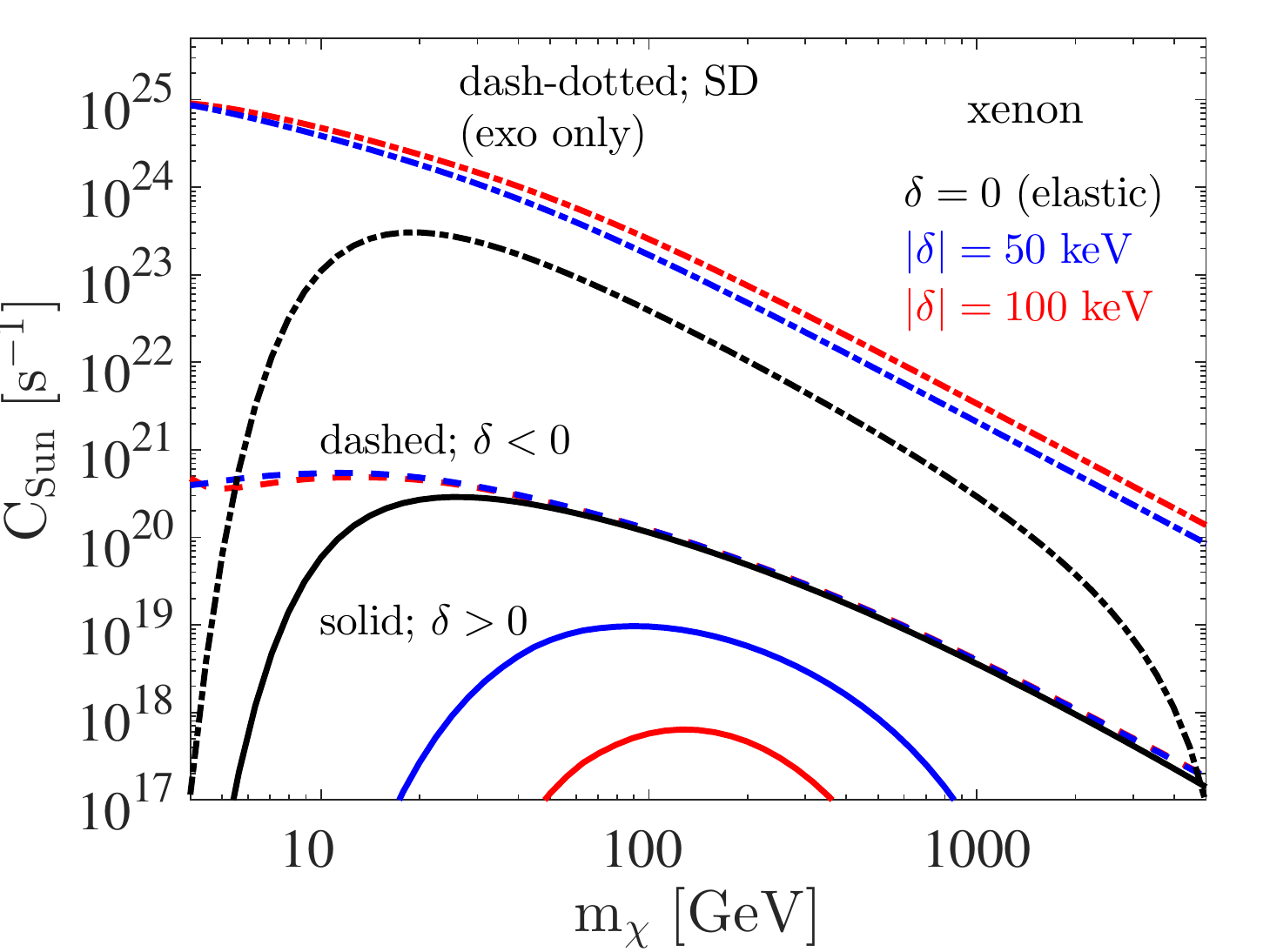}
	\caption{The lower bound on the capture rate as a function of the (true) DM mass for a xenon experiment. The blue (red) lines indicate $|\delta|=50\,(100)$~keV. The SI capture rates are shown for endothermic (solid) and exothermic (dashed) interactions. Also shown is the SD capture for exothermic DM as dash-dotted lines.} \label{bounds}
\end{figure}

The bounds using the generated mock data are shown for xenon in fig.~\ref{bounds} (left) as a function of the true DM mass. The SD capture rates are plotted using dash-dotted lines and the SI exothermic (endothermic)  with dashed (solid) lines. The colors indicate the absolute value of $\delta$ where red implies $|\delta|=100$~keV, blue $|\delta|=50$~keV and black refers to the elastic case. This would correspond to the case in which the DM is measured by some other means, such as the observation of a gamma ray line or via the combination of different DD signals~\cite{Kavanagh:2013wba}. As expected, the obtained bounds are the strongest for exothermic DM as the DD experiments probe the full velocity distribution, and thus the lower bound on the capture coincides with the actual capture rate. The capture rate for endothermic models is suppressed with respect to the elastic case since the velocity distribution is only probed for velocities where the SHM velocity distribution is suppressed.

\subsection{Combination with limits from neutrino observatories}
By assuming equilibrium, see eq.~\eqref{equilibrium2}, we may now combine these lower bounds with the upper bounds on the annihilation rate placed by neutrino telescopes. Data from SK~\cite{Choi:2015ara} is reported for DM masses in the range $4-200$~GeV for various channels (table I). We will use the results reported for annihilation into $b\bar{b}$ and $\tau \bar{\tau}$. The data is shown as upper bounds on the SD and SI cross sections, which we convert into upper limits on the capture rate and translate into upper limits on the annihilation rate by assuming equilibrium between capture and annihilation as per eq.~\eqref{equilibrium}. For larger masses, we use the results from ref.~\cite{Guo:2013ypa} (table II), which are based on data from an older SK search~\cite{Tanaka:2011uf}. We also use data for annihilation directly into $\nu_{\mu} \bar{\nu}_{\mu}$. For the IC results~\cite{Aartsen:2012kia} (table I), we use the data on the upper bound on the annihilation rate in the $b\bar{b}$ and $WW + \tau \bar{\tau}$ channels for DM masses in the range $20-5000$~GeV. In all cases, a branching ratio of 100\% into the specified channels is assumed. The direct consequence of the lower bound from a DD experiment being larger than the upper bound from neutrino telescopes is an upper bound on the branching ratios into these channels (see also ref.~\cite{Bernal:2012qh,Rott:2012qb,Rott:2015nma} for annihilations into MeV neutrinos).

\begin{figure} \label{bound_comparison_Xe}
	\centering
	\includegraphics[width=0.49\textwidth]{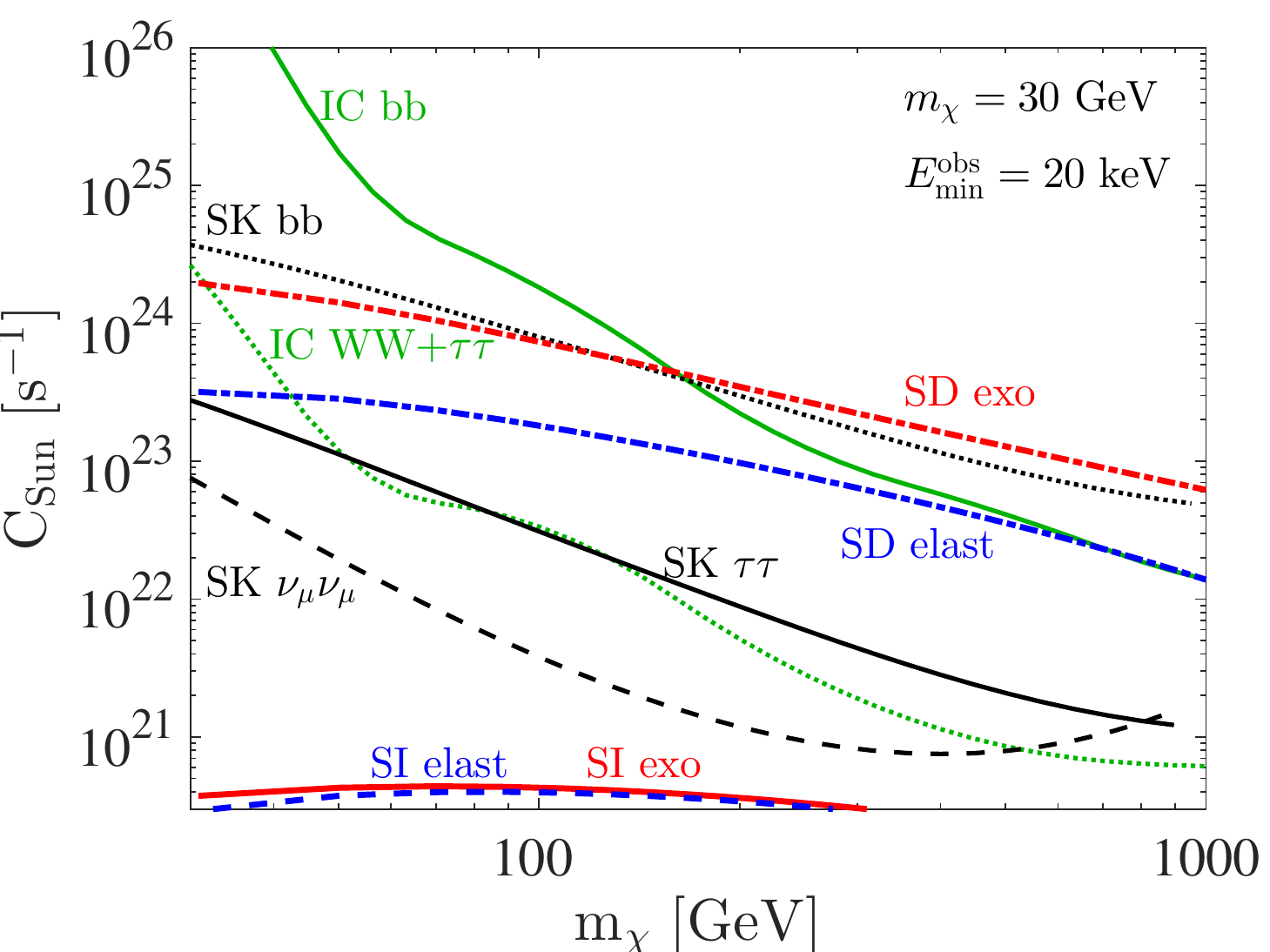}~~
	\includegraphics[width=0.49\textwidth]{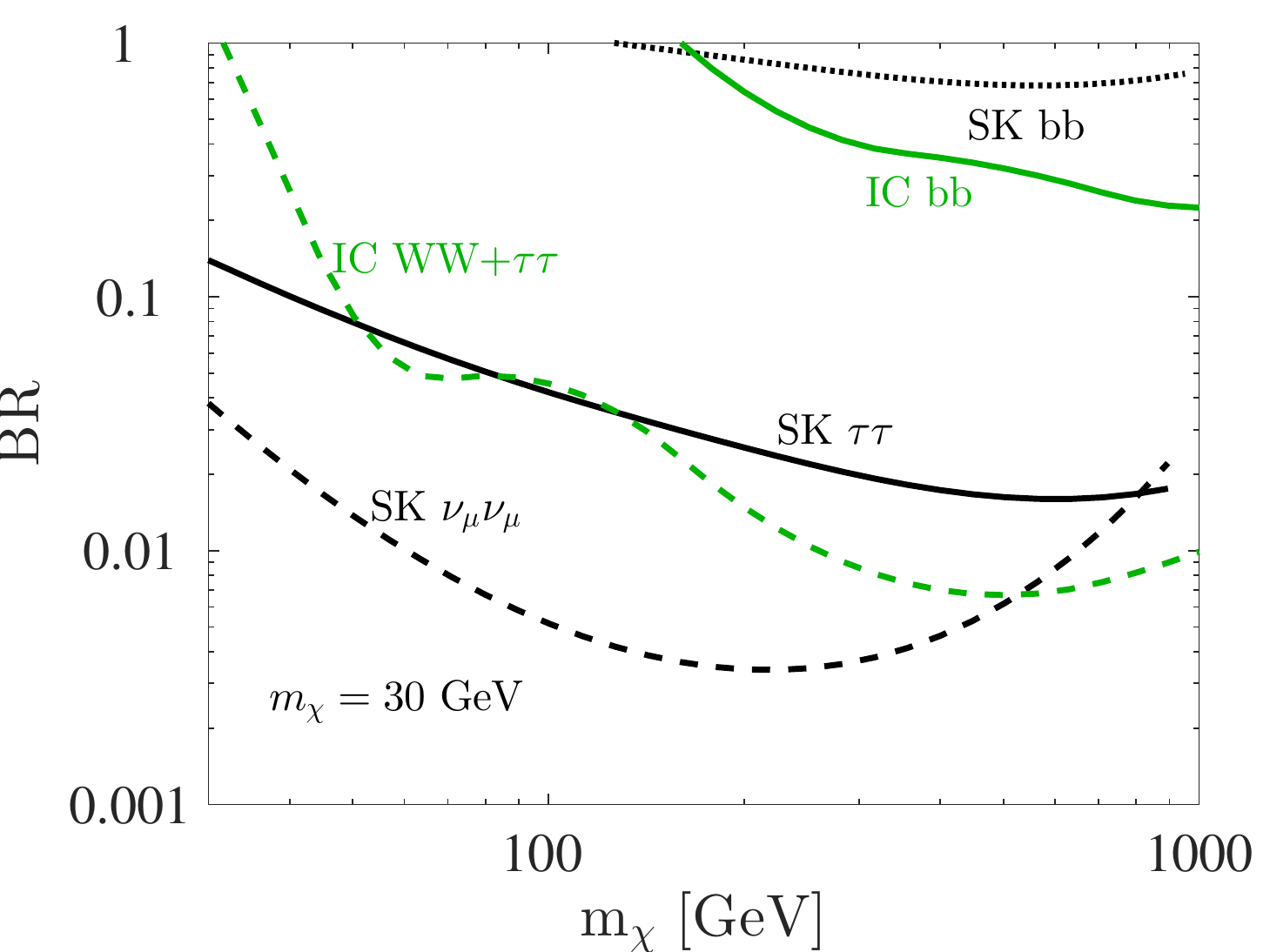}\\
	\includegraphics[width=0.49\textwidth]{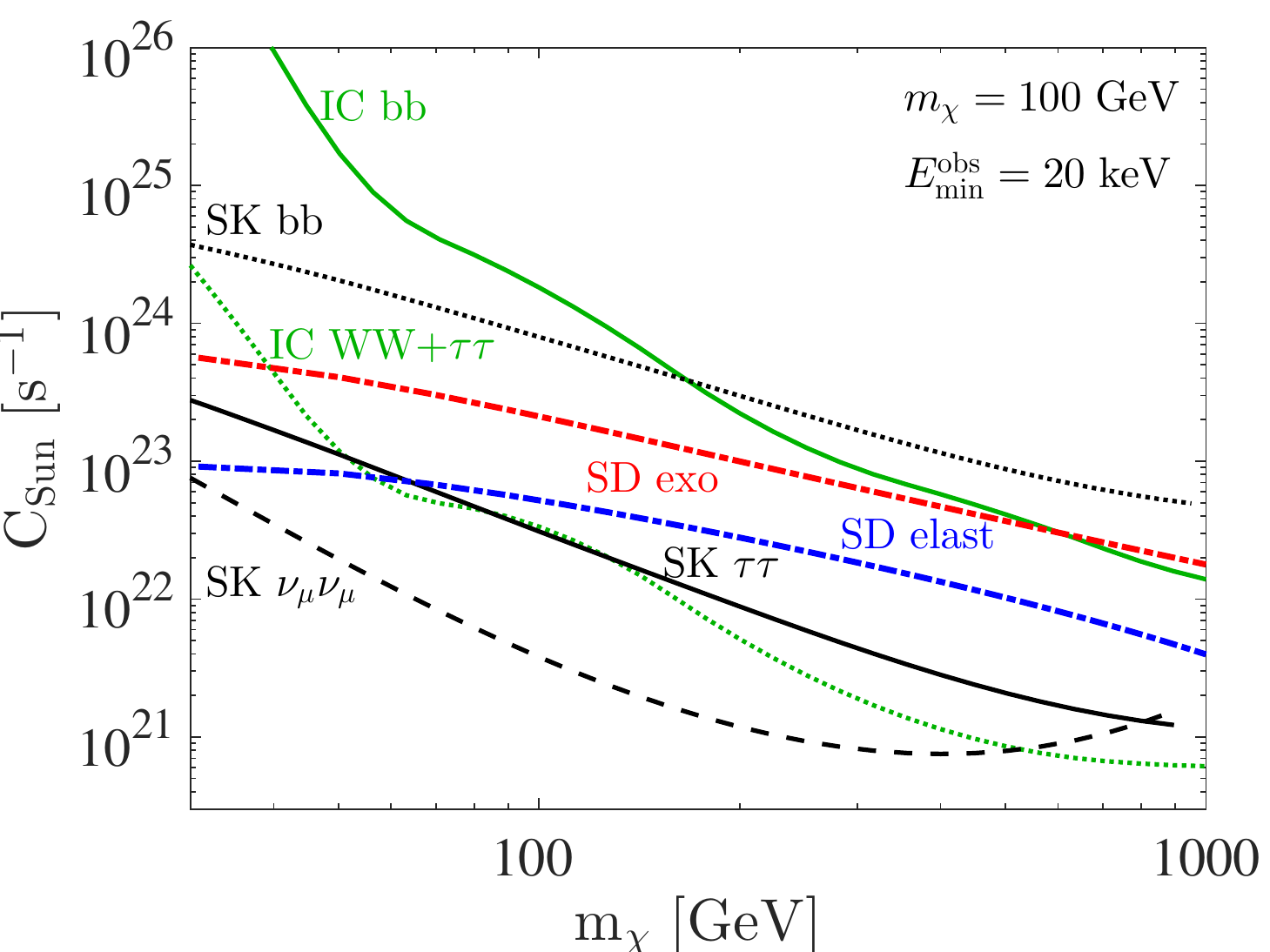}~~
	\includegraphics[width=0.49\textwidth]{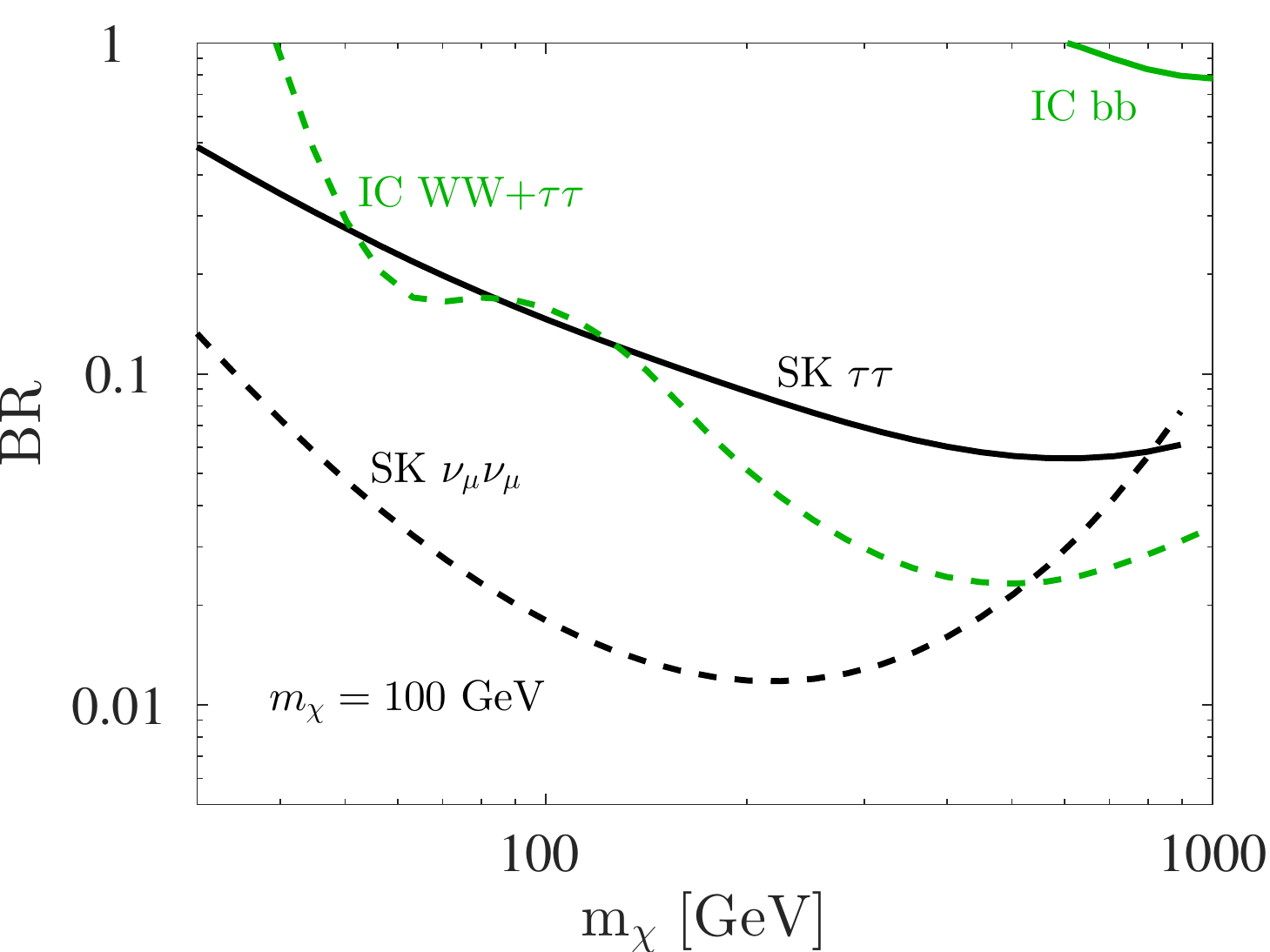}\\
	\includegraphics[width=0.49\textwidth]{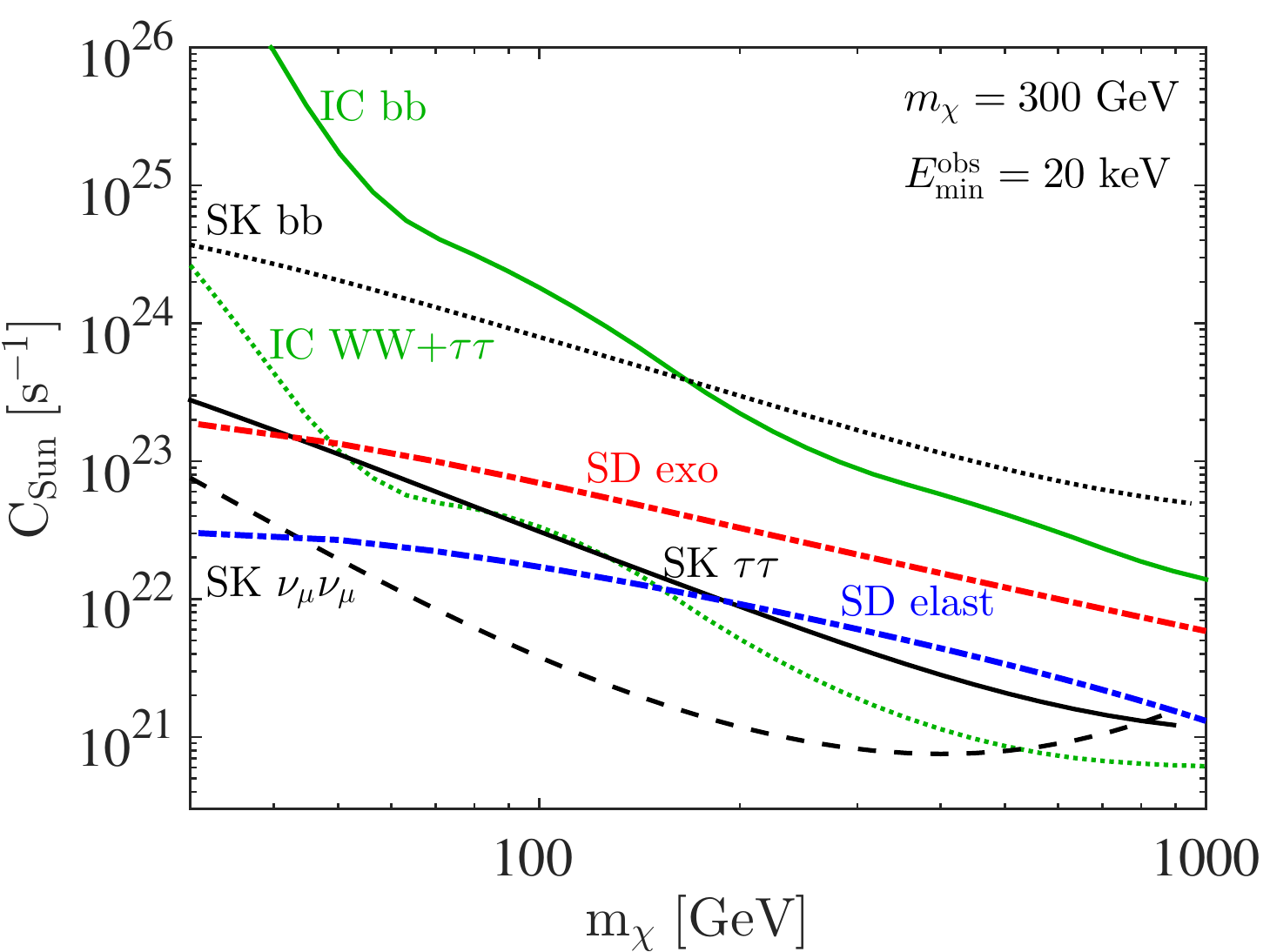}~~
	\includegraphics[width=0.49\textwidth]{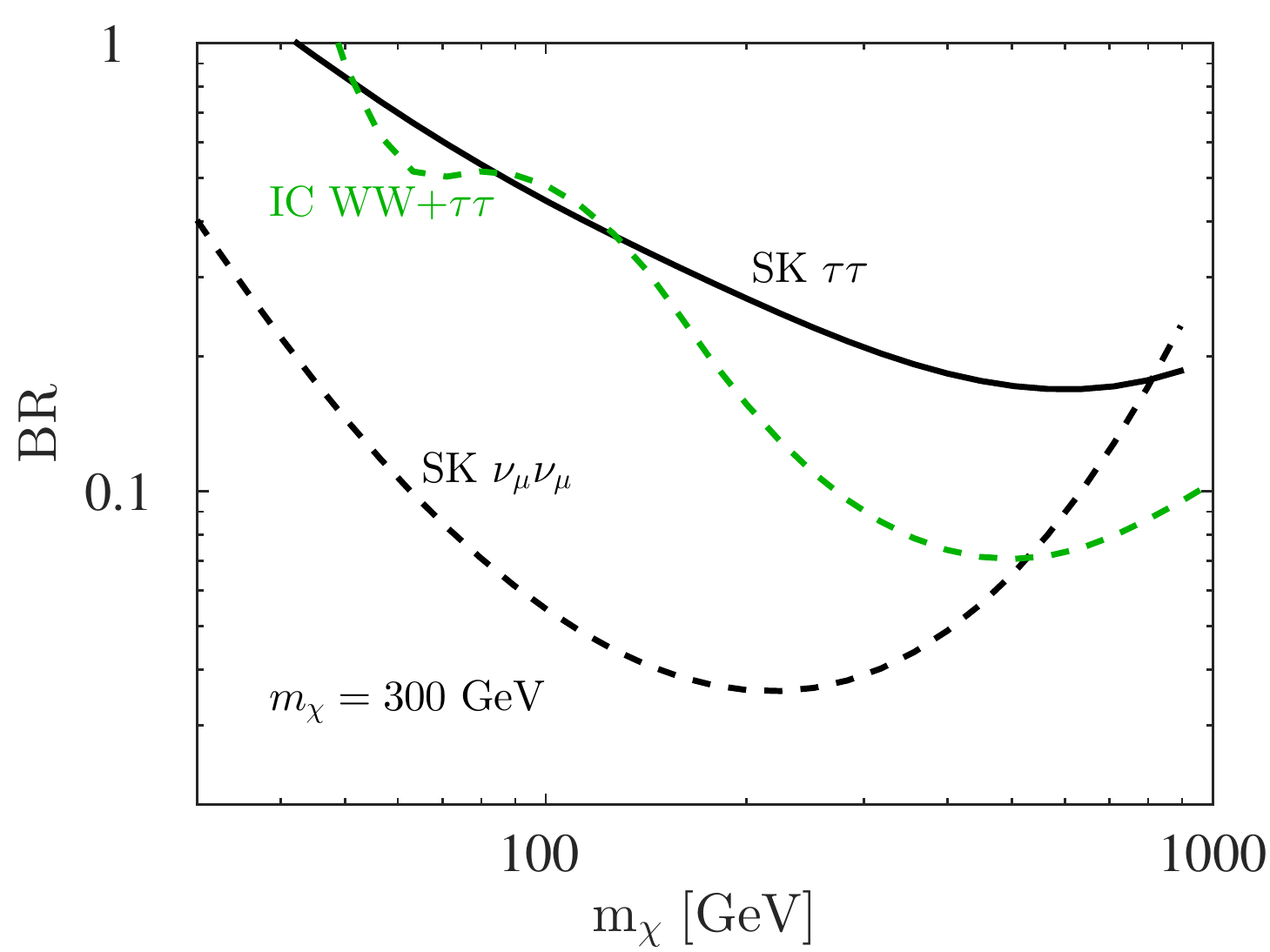}\\
	\caption{Left)  The lower bounds in the capture from a signal in a xenon experiment with $E^{\rm obs}_{\rm min}=20$~keV (see text) compared to limits from annihilation into the channels $\tau \tau$, $bb$ and $\nu_{\mu} \nu_{\mu}$ for SK, $\tau \tau + WW$ and $bb$ for IceCube, for a DM of $30$, $100$ and $300$~GeV in the upper, middle and bottom panels, respectively. Right) Upper bounds on the different channels for SD exothermic interactions.}
\end{figure}

To illustrate the strength of the bounds, we compute mock data for a DM particle with different (true) masses. We used $\sigma_{\rm SI}=10^{-45}$~cm$^2$ in the SI case and $\sigma_{\rm SD}=2\cdot 10^{-40}$~cm$^2$ in the SD case.\footnote{Notice that the upper bounds for endothermic (exothermic) are weaker (stronger), see figure \ref{vminxe}, and the precise upper bound depends on the DM mass.} Since the true DM mass is unknown, the bounds are computed with a mass different from the true one (which still enters in $\mathcal{C}$). The mass-splitting $\delta$ is also unknown but it can be related through the observed $E^{\rm obs}_{\rm min}$ and the DM mass by eq.~\eqref{dm}. The bounds resulting from a signal with a maximum of $\tilde{\eta}(E_R)$ at $E^{\rm obs}_{\rm min}=20$~keV are shown in fig.~\ref{bound_comparison_Xe} with the true DM mass of $30$, $100$ and $300$~GeV in the upper, middle and bottom plots, respectively. The mass range is chosen such that $\delta < 100$~keV is guaranteed. 

The right column shows the upper bounds on the branching ratios of the annihilation process into the specific channels. For SI interactions, no bounds can be placed (in the isospin-conserving case) unless $\delta$ is small and the DM mass is small enough for constraints on the SI cross section from DD experiments to become weak. Even if this is the case, the bounds on the branching ratios are weak. On the other hand, the bounds for the SD case can be very strong. For a true DM mass of $30$, $100$ and $300$~GeV respectively, the $\nu_{\mu} \nu_{\mu}$ channel is constrained below the 10\% level (for the $30$~GeV case,~1\%) in a large part of the scanned mass range. The SK~$\tau \tau$ and IC~$WW+\tau \tau$ channels are both constrained for all three cases, also in particular for the case of a $30$~GeV  DM particle. The bounds on the $bb$ branching ratios from both SK and IC are weak and occur only for the smaller DM masses when the assumed DM mass is much larger.

\begin{figure}
	\centering
	\includegraphics[width=0.6\textwidth]{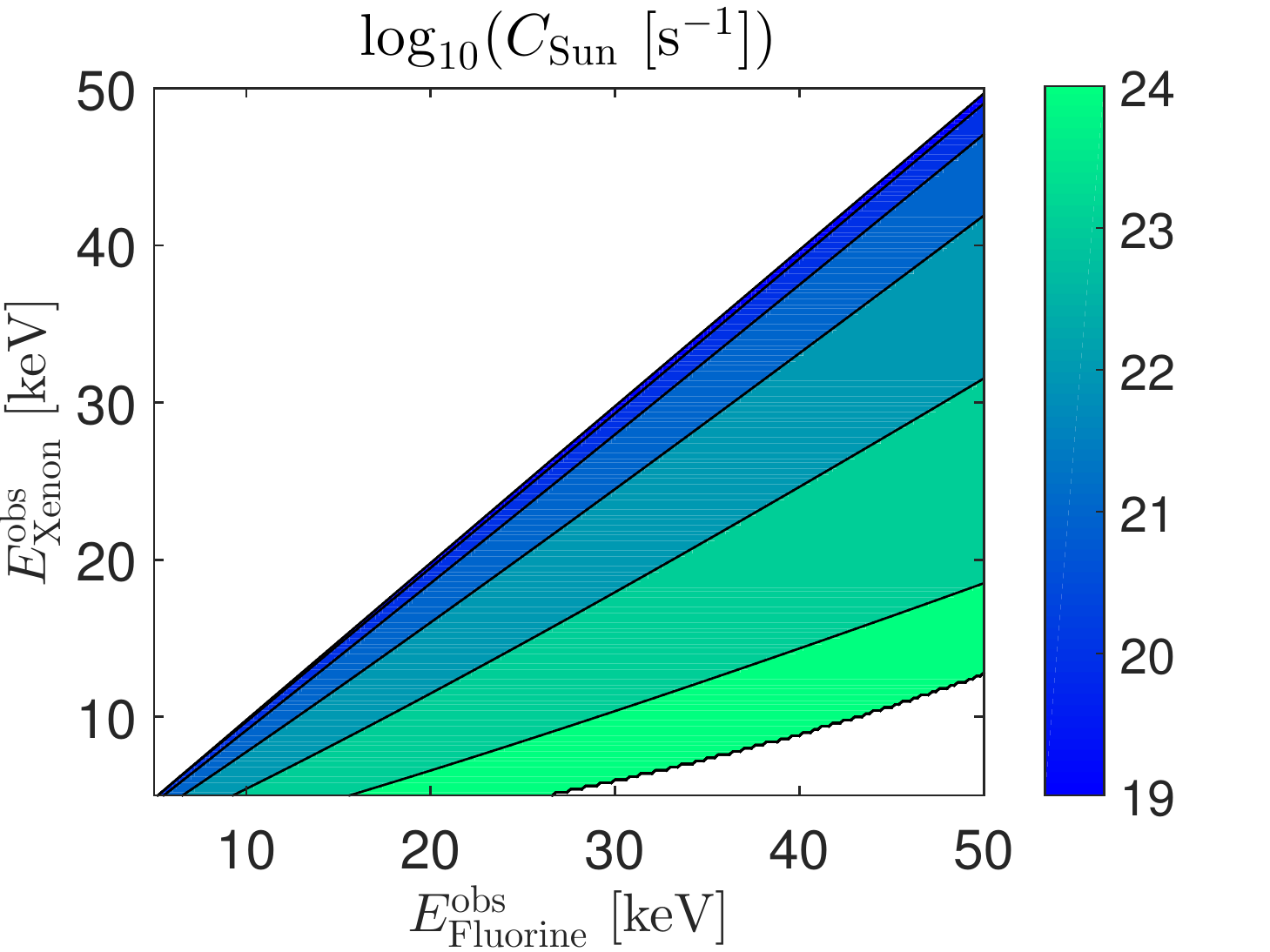}
	\caption{Density plot of the (logarithm of the) lower bounds on the capture rate in the sun, shown in the plane $E^{\rm obs}_{\rm Xe}- E^{\rm obs}_{\rm F}$, assuming SD exothermic interactions.} \label{2sig_exo}
\end{figure}

If a signal is observed in two DD experiments with different elements, one can extract the DM mass and the splitting of the excited states from the energy $E^{\rm obs}_{\rm min}$ at which the $\eta(E_R)$ is maximum, as explained in sec.~\ref{2signals}. 
This implies that one would have a precise determination of the lower bound on the capture rate, and thus on the neutrino flux for the different annihilation channels. In figure~\ref{2sig_exo}, we show the lower bound on the capture rate in the plane of the observed energies $E^{\rm obs}_{\rm min}$, see eq.~\eqref{minvmin}, in a fluorine (x-axis) and a xenon (y-axis) experiment, for the case of SD exothermic interactions. One can see that $E^{\rm obs}_{\rm Xe}<E^{\rm obs}_{\rm F}$. 

For a fixed energy, both the splitting $|\delta|$ and the DM mass grows with the mass of detector nuclei, see eq.~\eqref{minvmin}. In the case of exothermic interactions, this increase in the splitting for heavier nuclei is the dominant factor, and thus one can see that a large capture rate $\gtrsim 10^{22}\,{\rm s^{-1}}$ can be obtained for any observed $E^{\rm obs}_{\rm min}$ in xenon (larger $|\delta|$), while in fluorine $E^{\rm obs}_{\rm min}$ needs to be below $\sim 30$ keV (smaller $|\delta|$). The diagonal line corresponds to the case where the DM mass is very large, and thus the capture rate is suppressed. We have also checked the results for endothermic interactions, for which one finds closed contours, due the fact that in this case both large splittings and large DM masses suppressions go in the same direction, reducing the capture rates. In this case one can choose the strongest lower bounds of the two, which will be the one corresponding to the experiment that uses the heaviest element, and thus smallest $(v_{\rm m})_{\rm min}$, see eq.~\eqref{minvmin}.

From such a plot, once a DD signal is observed, one can infer the preferred energy range for another DD experiment to look for it, and the sensitivity that neutrino telescopes need to achieve in order to be able to also have a detection, which will of course depend on the annihilation channel. Note that in all our analysis we are conservative in the sense that we use mock data motivated by the future expected sensitivity in DD experiments combined with current upper limits from neutrino telescopes. It is certainly expected that if a DD signal is observed in next-generation of experiments, the upper limits from neutrino observatories will also be more stringent than current ones~\cite{Aartsen:2014oha, Coniglione:2015aqa, Abe:2011ts}.

\subsection{Isospin violation}
\begin{figure}
	\centering
	\includegraphics[width=0.6\textwidth]{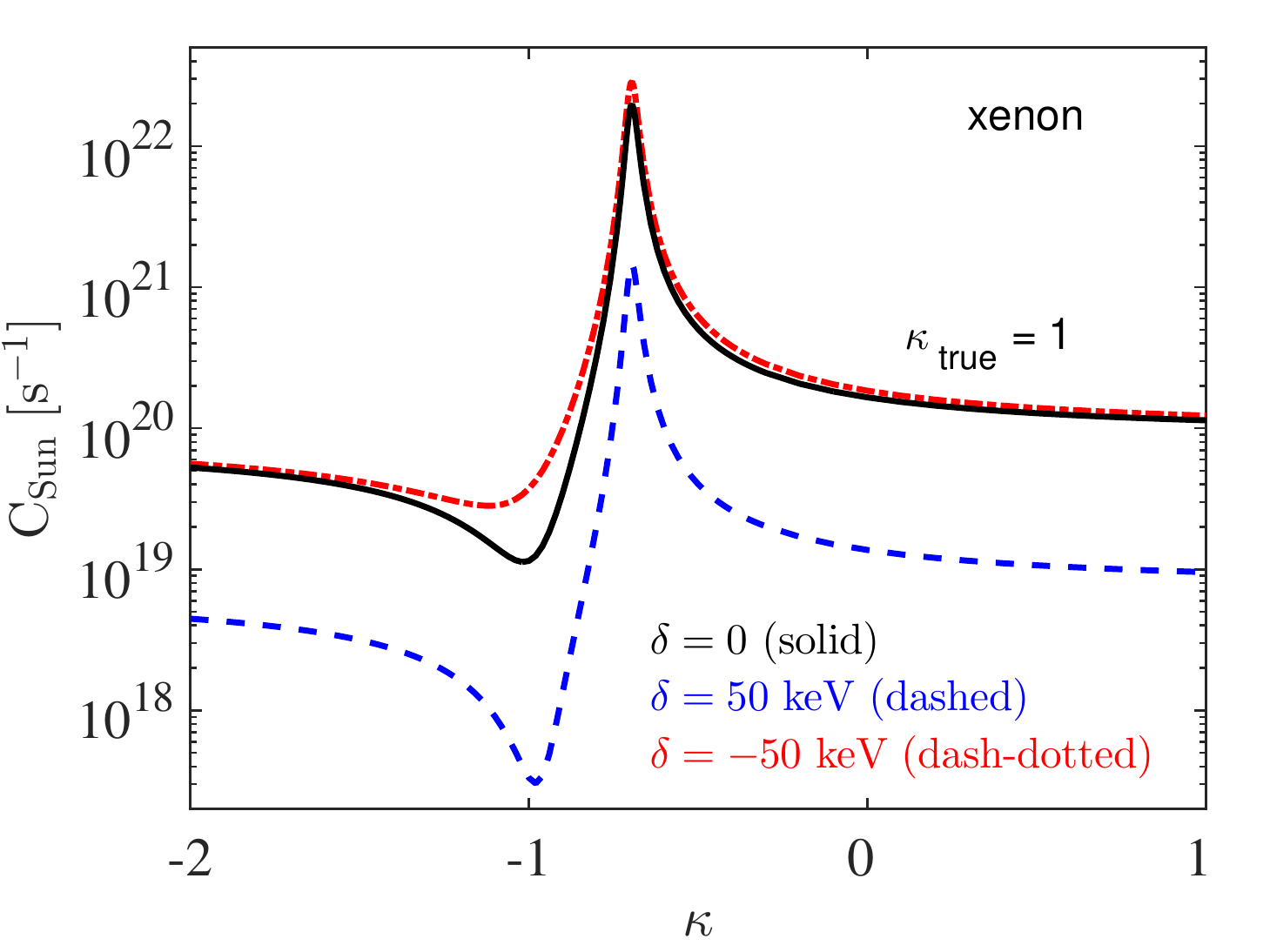}~~
	\caption{The lower bounds on the capture rate as a function of $\kappa=f_n/f_p$. We use $\kappa_{\rm true}=1$, $m_{\chi}=100$~GeV, a SI cross section of $10^{-45}$~cm$^2$ and $|\delta|=50$~keV. We also show  the elastic case.} \label{ISVcap_plot}
\end{figure}

Until now we have assumed that the couplings to the neutrons and the protons were equal and known in order to extract the velocity distribution, by using eq.~\eqref{extr_veldist}. The only unknown relevant parameter was the DM mass, as the splitting was given in terms of it by equation~\eqref{dm}, and thus the analysis was performed in terms of the DM mass only, see figure~\ref{bound_comparison_Xe}. However, for isospin-violating DM, the scattering rate due to a SI cross section will depend on $A_{\rm eff}^2 = \left( Z+\kappa\,(A-Z) \right)^2$, where $\kappa$ is the ratio of the DM coupling to the neutron and the proton, $\kappa=f_n/f_p$. If this is not the case and $\kappa$ is unknown, the extracted velocity distribution will be given by
\be
\mathcal{C} \tilde{f}_{\rm extr}(v) = \mathcal{C} \tilde{f}(v) \frac{A^2_{\rm true}}{A^2_{\rm eff}}\,,
\ee
where $A^2_{\rm true}$ is the value of $A^2_{\rm eff}$ for which $\kappa=\kappa_{\rm true}$, the true physical value. Here we assumed a perfect knowledge of the form factors (see ref.~\cite{Blennow:2015oea} for a similar analysis regarding the SD case, where their uncertainties are crucial). The impact of choosing the \emph{wrong} $\kappa$ is shown in fig.~\ref{ISVcap_plot}. Since the most abundant elements in the Sun have roughly the same number of protons and neutrons, destructive interference in the capture rate is expected for $\kappa = -1$. This will lower the expected capture rate at this choice of $\kappa$, as can be seen in the figure. However, if $\kappa$ is too close to the value which implies maximal destructive interference in the $A_{\rm eff}$ relevant for a DD experiment, then $\sigma_{\chi p}$ must be large in order to explain the presence of the DD signal in the first place, and thus the capture rate will be large. This is what is shown in the figure, where the peak in the capture rate for the signal in a xenon experiment occurs at $\kappa \sim -0.7$, where the destructive interference is the largest for this element. Thus the presence of isospin-violating couplings could imply larger lower bounds on the solar captures than expected, even in the endothermic case.

We discussed in sec.~\ref{2signals} the ability to extract the DM mass and the magnitude of two DD detection signals. In the case that DM interactions are isospin-violating, the only factor that change the extracted velocity distribution in two experiments is the coefficient $1/A_{\rm eff}^2$. The relative strength of the two signals will then be constant and one can relate $A_{\rm eff}^2$ in one experiment to the other. This allows for a determination of the isospin-violating parameter $\kappa$.

\section{Discussion and conclusions} \label{s:conc}

In this paper we have studied the capture of inelastic DM in the Sun for models with mass-splittings $\delta \leq 100$~keV, for both endothermic and exothermic scatterings with nuclei. We have extended the method of ref.~\cite{Blennow:2015oea}, which provides a lower bound on the capture rates in the Sun from a positive signal in a DD experiment, to include the case of inelastic DM. A single DD signal is not sufficient to separate exo- and endothermic interactions, and we have therefore treated both cases separately. Assuming that the excited state is stable and its abundance is similar to that of the lower mass state, the detection of the higher mass scattering in a DD experiment is much more likely. Of course, in this case, the lower bound on the capture will be given by that of the heavier state. If one assumed that the heavier state decays promptly, a signal will be from just endothermic interactions and so will the lower bound on the capture.

For the endothermic case, DD experiments only provide information on the high velocity range of the velocity distribution while exothermic scattering will make it possible to extract information on the full velocity distribution. The reason for this is the energy injection by down-scattering providing an energetic recoil in the experiment even at zero relative velocity between the DM and the target nuclei. This implies that the lower bounds on the capture rate are always larger for exothermic interactions than for endothermic ones. Furthermore, the bounds on the SD cross sections are weaker than for SI cross sections, and thus the capture rates for exothermic DM with primarily SD cross sections may be large, while no capture rate on hydrogen is possible for endothermic DM.

We have also discussed the implications of observing two different inelastic signals. First, from a measurement of both energies at which there is a maximum in the velocity integral one can extract both the dark matter mass and the mass splitting. Second, by going into velocity space and comparing the velocity integrals, one can test whether the scatterings are due to endothermic or exothermic DM interactions. A precise lower bound on the capture rate (and thus on the neutrino signal if equilibrium is assumed) can then be obtained in this case.

In order to perform the analysis, there are some assumptions regarding the particle physics model underneath the inelastic signal that need to be made. First, whether the heavier mass state is stable (exothermic and endothermic DM) or not (only endothermic DM). Second, that there is an elastic component in order for thermalization to occur. Third, that the annihilation cross section is such that equilibrium between the capture and the annihilation rate is reached.

To illustrate an application of the bounds presented here, we generated mock data for a future xenon experiment with a threshold energy of $3$~keV with SI and SD cross sections currently allowed by DD limits but close to the expected future sensitivity. We assume that a maximum of $\eta(E_R)$ will be observed at some energy $E^{\rm obs}_{\rm min}$. Using upper bounds from neutrino telescopes, the lower bounds on the capture rates can be converted into upper bounds on the branching ratios on the different annihilation channels. As expected, in the SI case the bounds are weak due to the small cross section unless isospin violation is present. For exothermic SD interactions, the capture rates may be very large and provide strong bounds on the branching ratios for DM annihilation into the $WW$, $\tau \tau$ and $\nu_\mu \nu_\mu$ channels.

Notice that in most models, where the cross section (at zero velocity) is the same for exothermic and endothermic scatterings, if a signal in a DD experiment is identified as due to exothermic scatterings, the extracted $\mathcal{C}\tilde{f}(v)$ in the whole velocity range, see eq.~\eqref{extr_veldist}, can be also used to compute the endothermic contribution to the DM capture rate. The lower bound on the capture rate is thus the sum of both the exothermic and the endothermic captures, as always $\rho_{\chi} \geq \rho_{\chi^*}$. This will strengthen the lower bound on the capture rate, since the endothermic capture rate is comparable to the exothermic one in a large region of the parameter space, c.f. figure~\ref{caprates}. In this sense, our final limits for exothermic scattering are very conservative.

In this paper we have focused on constant rates and $1/v^2$-dependent differential cross sections. In ref.~\cite{Blennow:2015oea} a lower bound valid for annual modulations~\cite{Drukier:1986tm, Freese:1987wu} in the case of elastic interactions was discussed, using the halo-independent bounds of refs.~\cite{HerreroGarcia:2011aa, HerreroGarcia:2012fu, Herrero-Garcia:2015kga}. The method described in Appendix A of ref.~\cite{Blennow:2015oea} can be combined with the framework discussed in this paper to test annual modulations in the case of inelastic scatterings. Moreover, the method described here can also be generalized to interactions that have cross sections in which the dependence on velocity and recoil energy factorize, see Appendix B of ref.~\cite{Blennow:2015oea}.

To conclude we want to emphasize that there are two effects that may alter the annihilation rate of captured DM particles that were not fully taken into account in the present analysis. First, we assumed a small elastic cross section to bring captured DM into a thermal distribution, which is expected in generic models of inelastic DM. However, one should keep in mind that a non-thermal distribution may decrease the annihilation rate. Second, if the heavier state does not immediately decay, there is an evaporation effect in which captured DM that have only had time to scatter a few times may down-scatter in the outer regions in the Sun. If the DM particle retains enough energy from the down-scatter, it may escape. This is partially avoided in our set-up by considering large enough DM masses. If a signal compatible with inelastic DM was ever observed, both effects should be resolved by a detailed study of the thermalization process of captured inelastic DM in the Sun.

\bigskip

{\bf Acknowledgements:} We are grateful to Sergio Palomares and Thomas Schwetz for discussions and comments on the manuscript. We also want to thank Sofia Sivertsson for discussions. This work was supported by the G{\"o}ran Gustafsson Foundation.

\bibliographystyle{my-h-physrev}
\bibliography{inelastic}

\begin{thebibliography}{10}

\bibitem{Lee.:2007qn}
KIMS, H.~S. Lee {\em et~al.},
\newblock {\em {Limits on WIMP-nucleon cross section with CsI(Tl) crystal
  detectors}},
\newblock Phys. Rev. Lett. {\bf 99}, 091301 (2007), 0704.0423.

\bibitem{Archambault:2009sm}
S.~Archambault {\em et~al.},
\newblock {\em {Dark Matter Spin-Dependent Limits for WIMP Interactions on F-19
  by PICASSO}},
\newblock Phys. Lett. {\bf B682}, 185 (2009), 0907.0307.

\bibitem{Bernabei:2010mq}
DAMA, LIBRA, R.~Bernabei {\em et~al.},
\newblock {\em {New results from DAMA/LIBRA}},
\newblock Eur. Phys. J. {\bf C67}, 39 (2010), 1002.1028.

\bibitem{Felizardo:2011uw}
M.~Felizardo {\em et~al.},
\newblock {\em {Final Analysis and Results of the Phase II SIMPLE Dark Matter
  Search}},
\newblock Phys. Rev. Lett. {\bf 108}, 201302 (2012), 1106.3014.

\bibitem{Armengaud:2012pfa}
EDELWEISS, E.~Armengaud {\em et~al.},
\newblock {\em {A search for low-mass WIMPs with EDELWEISS-II
  heat-and-ionization detectors}},
\newblock Phys. Rev. {\bf D86}, 051701 (2012), 1207.1815.

\bibitem{Aprile:2012nq}
XENON100, E.~Aprile {\em et~al.},
\newblock {\em {Dark Matter Results from 225 Live Days of XENON100 Data}},
\newblock Phys. Rev. Lett. {\bf 109}, 181301 (2012), 1207.5988.

\bibitem{Aalseth:2012if}
CoGeNT, C.~E. Aalseth {\em et~al.},
\newblock {\em {CoGeNT: A Search for Low-Mass Dark Matter using p-type Point
  Contact Germanium Detectors}},
\newblock Phys. Rev. {\bf D88}, 012002 (2013), 1208.5737.

\bibitem{Aprile:2013doa}
XENON100, E.~Aprile {\em et~al.},
\newblock {\em {Limits on spin-dependent WIMP-nucleon cross sections from 225
  live days of XENON100 data}},
\newblock Phys. Rev. Lett. {\bf 111}, 021301 (2013), 1301.6620.

\bibitem{Li:2013fla}
TEXONO, H.~B. Li {\em et~al.},
\newblock {\em {Limits on spin-independent couplings of WIMP dark matter with a
  p-type point-contact germanium detector}},
\newblock Phys. Rev. Lett. {\bf 110}, 261301 (2013), 1303.0925.

\bibitem{Agnese:2014aze}
SuperCDMS, R.~Agnese {\em et~al.},
\newblock {\em {Search for Low-Mass Weakly Interacting Massive Particles with
  SuperCDMS}},
\newblock Phys. Rev. Lett. {\bf 112}, 241302 (2014), 1402.7137.

\bibitem{Angloher:2014myn}
CRESST-II, G.~Angloher {\em et~al.},
\newblock {\em {Results on low mass WIMPs using an upgraded CRESST-II
  detector}},
\newblock Eur. Phys. J. {\bf C74}, 3184 (2014), 1407.3146.

\bibitem{Akimov:2011tj}
D.~{\relax Yu}. Akimov {\em et~al.},
\newblock {\em {WIMP-nucleon cross-section results from the second science run
  of ZEPLIN-III}},
\newblock Phys. Lett. {\bf B709}, 14 (2014), 1110.4769.

\bibitem{Akerib:2013tjd}
LUX, D.~S. Akerib {\em et~al.},
\newblock {\em {First results from the LUX dark matter experiment at the
  Sanford Underground Research Facility}},
\newblock Phys. Rev. Lett. {\bf 112}, 091303 (2014), 1310.8214.

\bibitem{Xiao:2014xyn}
PandaX, M.~Xiao {\em et~al.},
\newblock {\em {First dark matter search results from the PandaX-I
  experiment}},
\newblock Sci. China Phys. Mech. Astron. {\bf 57}, 2024 (2014), 1408.5114.

\bibitem{Amole:2015lsj}
PICO, C.~Amole {\em et~al.},
\newblock {\em {Dark Matter Search Results from the PICO-2L C$_3$F$_8$ Bubble
  Chamber}},
\newblock Phys. Rev. Lett. {\bf 114}, 231302 (2015), 1503.00008.

\bibitem{Akerib:2016lao}
LUX, D.~S. Akerib {\em et~al.},
\newblock {\em {First spin-dependent WIMP-nucleon cross section limits from the
  LUX experiment}},
\newblock (2016), 1602.03489.

\bibitem{Tanaka:2011uf}
Super-Kamiokande, T.~Tanaka {\em et~al.},
\newblock {\em {An Indirect Search for WIMPs in the Sun using 3109.6 days of
  upward-going muons in Super-Kamiokande}},
\newblock Astrophys. J. {\bf 742}, 78 (2011), 1108.3384.

\bibitem{Aartsen:2012kia}
IceCube, M.~G. Aartsen {\em et~al.},
\newblock {\em {Search for dark matter annihilations in the Sun with the
  79-string IceCube detector}},
\newblock Phys. Rev. Lett. {\bf 110}, 131302 (2013), 1212.4097.

\bibitem{Adrian-Martinez:2013ayv}
ANTARES, S.~Adrian-Martinez {\em et~al.},
\newblock {\em {First results on dark matter annihilation in the Sun using the
  ANTARES neutrino telescope}},
\newblock JCAP {\bf 1311}, 032 (2013), 1302.6516.

\bibitem{Choi:2015ara}
Super-Kamiokande, K.~Choi {\em et~al.},
\newblock {\em {Search for neutrinos from annihilation of captured low-mass
  dark matter particles in the Sun by Super-Kamiokande}},
\newblock Phys. Rev. Lett. {\bf 114}, 141301 (2015), 1503.04858.

\bibitem{Blennow:2015oea}
M.~Blennow, J.~Herrero-Garcia, and T.~Schwetz,
\newblock {\em {A halo-independent lower bound on the dark matter capture rate
  in the Sun from a direct detection signal}},
\newblock JCAP {\bf 1505}, 036 (2015), 1502.03342.

\bibitem{TuckerSmith:2001hy}
D.~Tucker-Smith and N.~Weiner,
\newblock {\em {Inelastic dark matter}},
\newblock Phys. Rev. {\bf D64}, 043502 (2001), hep-ph/0101138.

\bibitem{TuckerSmith:2004jv}
D.~Tucker-Smith and N.~Weiner,
\newblock {\em {The Status of inelastic dark matter}},
\newblock Phys. Rev. {\bf D72}, 063509 (2005), hep-ph/0402065.

\bibitem{Schwetz:2011aa}
T.~Schwetz and J.~Zupan,
\newblock {\em Dark Matter attempts for CoGeNT and DAMA},
\newblock (2011), 1106.6241.

\bibitem{Graham:2010ca}
P.~W. Graham, R.~Harnik, S.~Rajendran, and P.~Saraswat,
\newblock {\em {Exothermic Dark Matter}},
\newblock Phys. Rev. {\bf D82}, 063512 (2010), 1004.0937.

\bibitem{Frandsen:2014ima}
M.~T. Frandsen and I.~M. Shoemaker,
\newblock {\em {Up-shot of inelastic down-scattering at CDMS-Si}},
\newblock Phys. Rev. {\bf D89}, 051701 (2014), 1401.0624.

\bibitem{Chen:2014tka}
N.~Chen {\em et~al.},
\newblock {\em {Exothermic isospin-violating dark matter after SuperCDMS and
  CDEX}},
\newblock Phys. Lett. {\bf B743}, 205 (2015), 1404.6043.

\bibitem{Batell:2009vb}
B.~Batell, M.~Pospelov, and A.~Ritz,
\newblock {\em {Direct Detection of Multi-component Secluded WIMPs}},
\newblock Phys. Rev. {\bf D79}, 115019 (2009), 0903.3396.

\bibitem{Kopp:2009qt}
J.~Kopp, T.~Schwetz, and J.~Zupan,
\newblock {\em {Global interpretation of direct Dark Matter searches after
  CDMS-II results}},
\newblock JCAP {\bf 1002}, 014 (2010), 0912.4264.

\bibitem{Pospelov:2007mp}
M.~Pospelov, A.~Ritz, and M.~B. Voloshin,
\newblock {\em {Secluded WIMP Dark Matter}},
\newblock Phys. Lett. {\bf B662}, 53 (2008), 0711.4866.

\bibitem{Press:1985ug}
W.~H. Press and D.~N. Spergel,
\newblock {\em {Capture by the sun of a galactic population of weakly
  interacting massive particles}},
\newblock Astrophys. J. {\bf 296}, 679 (1985).

\bibitem{Griest:1987yu}
K.~Griest and D.~Seckel,
\newblock {\em {Cosmic Asymmetry, Neutrinos and the Sun}},
\newblock Nucl. Phys. {\bf B283}, 681 (1987),
\newblock [Erratum: Nucl. Phys.B296,1034(1988)].

\bibitem{Gould:1987ir}
A.~Gould,
\newblock {\em {Resonant Enhancements in WIMP Capture by the Earth}},
\newblock Astrophys. J. {\bf 321}, 571 (1987).

\bibitem{Ullio:2000bv}
P.~Ullio, M.~Kamionkowski, and P.~Vogel,
\newblock {\em {Spin dependent WIMPs in DAMA?}},
\newblock JHEP {\bf 07}, 044 (2001), hep-ph/0010036.

\bibitem{Peter:2009mk}
A.~H.~G. Peter,
\newblock {\em {Dark matter in the solar system II: WIMP annihilation rates in
  the Sun}},
\newblock Phys. Rev. {\bf D79}, 103532 (2009), 0902.1347.

\bibitem{Serpico:2010ae}
P.~D. Serpico and G.~Bertone,
\newblock {\em {Astrophysical limitations to the identification of dark matter:
  indirect neutrino signals vis-a-vis direct detection recoil rates}},
\newblock Phys. Rev. {\bf D82}, 063505 (2010), 1006.3268.

\bibitem{Liang:2013dsa}
Z.-L. Liang and Y.-L. Wu,
\newblock {\em {Direct detection and solar capture of spin-dependent dark
  matter}},
\newblock Phys. Rev. {\bf D89}, 013010 (2014), 1308.5897.

\bibitem{Choi:2013eda}
K.~Choi, C.~Rott, and Y.~Itow,
\newblock {\em {Impact of the dark matter velocity distribution on capture
  rates in the Sun}},
\newblock JCAP {\bf 1405}, 049 (2014), 1312.0273.

\bibitem{Kavanagh:2014rya}
B.~J. Kavanagh, M.~Fornasa, and A.~M. Green,
\newblock {\em {Probing WIMP particle physics and astrophysics with direct
  detection and neutrino telescope data}},
\newblock Phys. Rev. {\bf D91}, 103533 (2015), 1410.8051.

\bibitem{Arina:2013jya}
C.~Arina, G.~Bertone, and H.~Silverwood,
\newblock {\em {Complementarity of direct and indirect Dark Matter detection
  experiments}},
\newblock Phys. Rev. {\bf D88}, 013002 (2013), 1304.5119.

\bibitem{Kappl:2011kz}
R.~Kappl and M.~W. Winkler,
\newblock {\em {New Limits on Dark Matter from Super-Kamiokande}},
\newblock Nucl. Phys. {\bf B850}, 505 (2011), 1104.0679.

\bibitem{Wikstrom:2009kw}
G.~Wikstr{\"o}m and J.~Edsj{\"o},
\newblock {\em {Limits on the WIMP-nucleon scattering cross-section from
  neutrino telescopes}},
\newblock JCAP {\bf 0904}, 009 (2009), 0903.2986.

\bibitem{Hooper:2008cf}
D.~Hooper, F.~Petriello, K.~M. Zurek, and M.~Kamionkowski,
\newblock {\em {The New DAMA Dark-Matter Window and Energetic-Neutrino
  Searches}},
\newblock Phys. Rev. {\bf D79}, 015010 (2009), 0808.2464.

\bibitem{Kamionkowski:1994dp}
M.~Kamionkowski, K.~Griest, G.~Jungman, and B.~Sadoulet,
\newblock {\em {Model independent comparison of direct versus indirect
  detection of supersymmetric dark matter}},
\newblock Phys. Rev. Lett. {\bf 74}, 5174 (1995), hep-ph/9412213.

\bibitem{Bergstrom:1998xh}
L.~Bergstr{\"o}m, J.~Edsj{\"o}, and P.~Gondolo,
\newblock {\em {Indirect detection of dark matter in km size neutrino
  telescopes}},
\newblock Phys. Rev. {\bf D58}, 103519 (1998), hep-ph/9806293.

\bibitem{Ibarra:2014vya}
A.~Ibarra, M.~Totzauer, and S.~Wild,
\newblock {\em {Higher order dark matter annihilations in the Sun and
  implications for IceCube}},
\newblock JCAP {\bf 1404}, 012 (2014), 1402.4375.

\bibitem{Catena:2015iea}
R.~Catena,
\newblock {\em {Dark matter signals at neutrino telescopes in effective
  theories}},
\newblock JCAP {\bf 1504}, 052 (2015), 1503.04109.

\bibitem{Nussinov:2009ft}
S.~Nussinov, L.-T. Wang, and I.~Yavin,
\newblock {\em {Capture of Inelastic Dark Matter in the Sun}},
\newblock JCAP {\bf 0908}, 037 (2009), 0905.1333.

\bibitem{Menon:2009qj}
A.~Menon, R.~Morris, A.~Pierce, and N.~Weiner,
\newblock {\em {Capture and Indirect Detection of Inelastic Dark Matter}},
\newblock Phys. Rev. {\bf D82}, 015011 (2010), 0905.1847.

\bibitem{Shu:2010ta}
J.~Shu, P.-f. Yin, and S.-h. Zhu,
\newblock {\em {Neutrino Constraints on Inelastic Dark Matter after CDMS II}},
\newblock Phys. Rev. {\bf D81}, 123519 (2010), 1001.1076.

\bibitem{Catena:2009mf}
R.~Catena and P.~Ullio,
\newblock {\em {A novel determination of the local dark matter density}},
\newblock JCAP {\bf 1008}, 004 (2010), 0907.0018.

\bibitem{Weber:2009pt}
M.~Weber and W.~de~Boer,
\newblock {\em {Determination of the Local Dark Matter Density in our Galaxy}},
\newblock Astron. Astrophys. {\bf 509}, A25 (2010), 0910.4272.

\bibitem{Salucci:2010qr}
P.~Salucci, F.~Nesti, G.~Gentile, and C.~F. Martins,
\newblock {\em {The dark matter density at the Sun's location}},
\newblock Astron. Astrophys. {\bf 523}, A83 (2010), 1003.3101.

\bibitem{Pato:2010yq}
M.~Pato, O.~Agertz, G.~Bertone, B.~Moore, and R.~Teyssier,
\newblock {\em {Systematic uncertainties in the determination of the local dark
  matter density}},
\newblock Phys. Rev. {\bf D82}, 023531 (2010), 1006.1322.

\bibitem{Iocco:2011jz}
F.~Iocco, M.~Pato, G.~Bertone, and P.~Jetzer,
\newblock {\em {Dark Matter distribution in the Milky Way: microlensing and
  dynamical constraints}},
\newblock JCAP {\bf 1111}, 029 (2011), 1107.5810.

\bibitem{Garbari:2012ff}
S.~Garbari, C.~Liu, J.~I. Read, and G.~Lake,
\newblock {\em {A new determination of the local dark matter density from the
  kinematics of K dwarfs}},
\newblock Mon. Not. Roy. Astron. Soc. {\bf 425}, 1445 (2012), 1206.0015.

\bibitem{Bovy:2012tw}
J.~Bovy and S.~Tremaine,
\newblock {\em {On the local dark matter density}},
\newblock Astrophys. J. {\bf 756}, 89 (2012), 1205.4033.

\bibitem{Cui:2009xq}
Y.~Cui, D.~E. Morrissey, D.~Poland, and L.~Randall,
\newblock {\em {Candidates for Inelastic Dark Matter}},
\newblock JHEP {\bf 05}, 076 (2009), 0901.0557.

\bibitem{Gould:1987ju}
A.~Gould,
\newblock {\em {{WIMP} Distribution in and Evaporation From the Sun}},
\newblock Astrophys. J. {\bf 321}, 560 (1987).

\bibitem{Busoni:2013kaa}
G.~Busoni, A.~De~Simone, and W.-C. Huang,
\newblock {\em {On the Minimum Dark Matter Mass Testable by Neutrinos from the
  Sun}},
\newblock JCAP {\bf 1307}, 010 (2013), 1305.1817.

\bibitem{Blennow:2007tw}
M.~Blennow, J.~Edsj{\"o}, and T.~Ohlsson,
\newblock {\em {Neutrinos from WIMP annihilations using a full three-flavor
  Monte Carlo}},
\newblock JCAP {\bf 0801}, 021 (2008), 0709.3898.

\bibitem{Cirelli:2005gh}
M.~Cirelli {\em et~al.},
\newblock {\em {Spectra of neutrinos from dark matter annihilations}},
\newblock Nucl. Phys. {\bf B727}, 99 (2005), hep-ph/0506298,
\newblock [Erratum: Nucl. Phys.B790,338(2008)].

\bibitem{Goodman:1984dc}
M.~W. Goodman and E.~Witten,
\newblock {\em {Detectability of Certain Dark Matter Candidates}},
\newblock Phys. Rev. {\bf D31}, 3059 (1985).

\bibitem{Drukier:1986tm}
A.~K. Drukier, K.~Freese, and D.~N. Spergel,
\newblock {\em {Detecting Cold Dark Matter Candidates}},
\newblock Phys. Rev. {\bf D33}, 3495 (1986).

\bibitem{Freese:1987wu}
K.~Freese, J.~A. Frieman, and A.~Gould,
\newblock {\em {Signal Modulation in Cold Dark Matter Detection}},
\newblock Phys. Rev. {\bf D37}, 3388 (1988).

\bibitem{HerreroGarcia:2011aa}
J.~Herrero-Garcia, T.~Schwetz, and J.~Zupan,
\newblock {\em {On the annual modulation signal in dark matter direct
  detection}},
\newblock JCAP {\bf 1203}, 005 (2012), 1112.1627.

\bibitem{HerreroGarcia:2012fu}
J.~Herrero-Garcia, T.~Schwetz, and J.~Zupan,
\newblock {\em {Astrophysics independent bounds on the annual modulation of
  dark matter signals}},
\newblock Phys. Rev. Lett. {\bf 109}, 141301 (2012), 1205.0134.

\bibitem{Herrero-Garcia:2015kga}
J.~Herrero-Garcia,
\newblock {\em {Halo-independent tests of dark matter annual modulation
  signals}},
\newblock JCAP {\bf 1509}, 012 (2015), 1506.03503.

\bibitem{Fox:2011bu}
P.~J. Fox, G.~D. Kribs, and T.~M.~P. Tait,
\newblock {\em {Interpreting Dark Matter Direct Detection Independently of the
  Local Velocity and Density Distribution}},
\newblock Phys. Rev. {\bf D83}, 034007 (2011), 1011.1910.

\bibitem{Fox:2011bz}
P.~J. Fox, J.~Liu, and N.~Weiner,
\newblock {\em {Integrating Out Astrophysical Uncertainties}},
\newblock Phys. Rev. {\bf D83}, 103514 (2011), 1011.1915.

\bibitem{Bozorgnia:2013hsa}
N.~Bozorgnia, J.~Herrero-Garcia, T.~Schwetz, and J.~Zupan,
\newblock {\em {Halo-independent methods for inelastic dark matter
  scattering}},
\newblock JCAP {\bf 1307}, 049 (2013), 1305.3575.

\bibitem{McCabe:2011sr}
C.~McCabe,
\newblock {\em {DAMA and CoGeNT without astrophysical uncertainties}},
\newblock Phys. Rev. {\bf D84}, 043525 (2011), 1107.0741.

\bibitem{McCabe:2010zh}
C.~McCabe,
\newblock {\em {The Astrophysical Uncertainties Of Dark Matter Direct Detection
  Experiments}},
\newblock Phys. Rev. {\bf D82}, 023530 (2010), 1005.0579.

\bibitem{Frandsen:2011gi}
M.~T. Frandsen, F.~Kahlhoefer, C.~McCabe, S.~Sarkar, and K.~Schmidt-Hoberg,
\newblock {\em {Resolving astrophysical uncertainties in dark matter direct
  detection}},
\newblock JCAP {\bf 1201}, 024 (2012), 1111.0292.

\bibitem{DelNobile:2013cta}
E.~Del~Nobile, G.~B. Gelmini, P.~Gondolo, and J.-H. Huh,
\newblock {\em {Halo-independent analysis of direct detection data for light
  WIMPs}},
\newblock JCAP {\bf 1310}, 026 (2013), 1304.6183.

\bibitem{DelNobile:2013cva}
E.~Del~Nobile, G.~Gelmini, P.~Gondolo, and J.-H. Huh,
\newblock {\em {Generalized Halo Independent Comparison of Direct Dark Matter
  Detection Data}},
\newblock JCAP {\bf 1310}, 048 (2013), 1306.5273.

\bibitem{Bozorgnia:2014gsa}
N.~Bozorgnia and T.~Schwetz,
\newblock {\em {What is the probability that direct detection experiments have
  observed Dark Matter?}},
\newblock JCAP {\bf 1412}, 015 (2014), 1410.6160.

\bibitem{Fox:2014kua}
P.~J. Fox, Y.~Kahn, and M.~McCullough,
\newblock {\em {Taking Halo-Independent Dark Matter Methods Out of the Bin}},
\newblock JCAP {\bf 1410}, 076 (2014), 1403.6830.

\bibitem{Feldstein:2014ufa}
B.~Feldstein and F.~Kahlhoefer,
\newblock {\em {Quantifying (dis)agreement between direct detection experiments
  in a halo-independent way}},
\newblock JCAP {\bf 1412}, 052 (2014), 1409.5446.

\bibitem{Cherry:2014wia}
J.~F. Cherry, M.~T. Frandsen, and I.~M. Shoemaker,
\newblock {\em {Halo Independent Direct Detection of Momentum-Dependent Dark
  Matter}},
\newblock JCAP {\bf 1410}, 022 (2014), 1405.1420.

\bibitem{Blennow:2015gta}
M.~Blennow, J.~Herrero-Garcia, T.~Schwetz, and S.~Vogl,
\newblock {\em {Halo-independent tests of dark matter direct detection signals:
  local DM density, LHC, and thermal freeze-out}},
\newblock JCAP {\bf 1508}, 039 (2015), 1505.05710.

\bibitem{Ferrer:2015bta}
F.~Ferrer, A.~Ibarra, and S.~Wild,
\newblock {\em {A novel approach to derive halo-independent limits on dark
  matter properties}},
\newblock JCAP {\bf 1509}, 052 (2015), 1506.03386.

\bibitem{Drees:2007hr}
M.~Drees and C.-L. Shan,
\newblock {\em {Reconstructing the Velocity Distribution of WIMPs from Direct
  Dark Matter Detection Data}},
\newblock JCAP {\bf 0706}, 011 (2007), astro-ph/0703651.

\bibitem{Drees:2008bv}
M.~Drees and C.-L. Shan,
\newblock {\em {Model-Independent Determination of the WIMP Mass from Direct
  Dark Matter Detection Data}},
\newblock JCAP {\bf 0806}, 012 (2008), 0803.4477.

\bibitem{Kavanagh:2013wba}
B.~J. Kavanagh and A.~M. Green,
\newblock {\em {Model independent determination of the dark matter mass from
  direct detection experiments}},
\newblock Phys. Rev. Lett. {\bf 111}, 031302 (2013), 1303.6868.

\bibitem{Malling:2011va}
D.~C. Malling {\em et~al.},
\newblock {\em {After LUX: The LZ Program}},
\newblock (2011), 1110.0103.

\bibitem{Baudis:2012bc}
DARWIN Consortium, L.~Baudis,
\newblock {\em {DARWIN: dark matter WIMP search with noble liquids}},
\newblock J. Phys. Conf. Ser. {\bf 375}, 012028 (2012), 1201.2402.

\bibitem{Aprile:2012zx}
XENON1T, E.~Aprile,
\newblock {\em {The XENON1T Dark Matter Search Experiment}},
\newblock Springer Proc. Phys. {\bf 148}, 93 (2013), 1206.6288.

\bibitem{Aprile:2015uzo}
XENON, E.~Aprile {\em et~al.},
\newblock {\em {Physics reach of the XENON1T dark matter experiment}},
\newblock Submitted to: JCAP  (2015), 1512.07501.

\bibitem{Serenelli:2009yc}
A.~Serenelli, S.~Basu, J.~W. Ferguson, and M.~Asplund,
\newblock {\em {New Solar Composition: The Problem With Solar Models
  Revisited}},
\newblock Astrophys. J. {\bf 705}, L123 (2009), 0909.2668.

\bibitem{Guo:2013ypa}
W.-L. Guo, Z.-L. Liang, and Y.-L. Wu,
\newblock {\em {Direct detection and solar capture of dark matter with momentum
  and velocity dependent elastic scattering}},
\newblock Nucl. Phys. {\bf B878}, 295 (2014), 1305.0912.

\bibitem{Bernal:2012qh}
N.~Bernal, J.~Mart{\'\i}n-Albo, and S.~Palomares-Ruiz,
\newblock {\em {A novel way of constraining WIMPs annihilations in the Sun: MeV
  neutrinos}},
\newblock JCAP {\bf 1308}, 011 (2013), 1208.0834.

\bibitem{Rott:2012qb}
C.~Rott, J.~Siegal-Gaskins, and J.~F. Beacom,
\newblock {\em {New Sensitivity to Solar WIMP Annihilation using Low-Energy
  Neutrinos}},
\newblock Phys. Rev. {\bf D88}, 055005 (2013), 1208.0827.

\bibitem{Rott:2015nma}
C.~Rott, S.~In, J.~Kumar, and D.~Yaylali,
\newblock {\em {Dark Matter Searches for Monoenergetic Neutrinos Arising from
  Stopped Meson Decay in the Sun}},
\newblock JCAP {\bf 1511}, 039 (2015), 1510.00170.

\bibitem{Aartsen:2014oha}
IceCube PINGU, M.~G. Aartsen {\em et~al.},
\newblock {\em {Letter of Intent: The Precision IceCube Next Generation Upgrade
  (PINGU)}},
\newblock (2014), 1401.2046.

\bibitem{Coniglione:2015aqa}
KM3NeT, R.~Coniglione,
\newblock {\em {The KM3NeT neutrino telescope}},
\newblock J. Phys. Conf. Ser. {\bf 632}, 012002 (2015).

\bibitem{Abe:2011ts}
K.~Abe {\em et~al.},
\newblock {\em {Letter of Intent: The Hyper-Kamiokande Experiment --- Detector
  Design and Physics Potential ---}},
\newblock (2011), 1109.3262.

\end{thebibliography}

\end{document}